\documentclass{aa}  
\usepackage{graphicx}
\usepackage[varg]{txfonts}
\usepackage{xcolor}
\usepackage{amsmath}
\usepackage{float}
\usepackage{sidecap}

\usepackage{natbib}
\bibpunct{(}{)}{;}{a}{}{,}

\usepackage[colorlinks=True,citecolor=blue,linkcolor=blue]{hyperref}

\newcommand{\vect}[1] {\mbox{\boldmath ${#1}$}}

\newcommand{\Lxmes}[0]{\hat{L_X}}
\newcommand{\Tmes}[0]{\hat{T}}
\newcommand{\zmes}[0]{\hat{z}}

\usepackage{amssymb}
\usepackage{amsmath}
\usepackage{mathrsfs}
\usepackage{mathtools}
\usepackage{calrsfs}

\begin{document} 

\defcitealias{2021Koulouridis}{K21}

   \title{Identification of low redshift groups and clusters of galaxies in the X-CLASS survey and the X-ray luminosity-temperature relation\thanks{Partly based on observations made at Observatoire de Haute Provence (CNRS), France, with MISTRAL on the T193 telescope.}}
   \titlerunning{Low redshift groups and clusters of galaxies in X-CLASS}

   \author{Q.~Moysan \inst{1}
         \and
         F.~Sarron \inst{1}
         \and
         N.~Clerc \inst{1}
         \and
         G.~Soucail \inst{1}
         \and
         C.~Adami \inst{2}
         \and
         B.~Altieri \inst{3}
         \and
         R.~Cabanac \inst{1}
         \and
         M.~Chira \inst{4}
         \and
         J.~Comparat \inst{5}
         \and
         D.~Coia \inst{6}
         \and
         E.~Drigga \inst{4}
         \and
         E.~Gaynullina \inst{7}
         \and
         A.~Khalikova \inst{7}
         \and
         E.~Koulouridis \inst{4,8}
         \and
         K.~Migkas \inst{9}
         \and
         M.~Molham \inst{10}
         \and
         L.~Paquereau \inst{11}
         \and
         T.~Sadibekova \inst{8}
         \and
         I.~Valtchanov \inst{6}
         }

   \institute{
        IRAP, Universit{\'e} de Toulouse, CNRS, CNES, 14 Avenue Belin, 31400 Toulouse, France \\
        \email{quentin.moysan@utoulouse.fr}
        \and
        LAM, Aix-Marseille Universit{\'e}, CNRS, CNES, 38, rue Fr{\'e}d{\'e}ric Joliot-Curie, 13388 Marseille Cedex 13, France 
        \and
        ESAC/ESA, Camino Bajo del Castillo, s/n., Urb. Villafranca del Castillo, 28692 Villanueva de la Ca{\~n}ada, Madrid, Spain
        \and
        Institute for Astronomy \& Astrophysics, National Observatory of Athens, 15236 Palaia Penteli, Greece
        \and
        Univ. Grenoble Alpes, CNRS, Grenoble INP, LPSC-IN2P3, 53, Avenue des Martyrs, 38000, Grenoble, France
        \and
        Telespazio UK for ESA, European Space Astronomy Centre, Operations Department, E-28691 Villanueva de la Ca{\~n}ada, Spain
        \and
        Ulugh Beg Astronomical Institute of Uzbekistan Academy of Sciences, 33 Astronomicheskaya Str., Tashkent 100052, Uzbekistan
        \and
        Universit{\'e} Paris-Saclay, Universit{\'e} Paris Cit{\'e}, CEA, CNRS, AIM, 91191, Gif-sur-Yvette, France
        \and
        Leiden Observatory, Leiden University, P.O. Box 9513, 2300 RA Leiden, The Netherlands
        \and
        National Research Institute of Astronomy and Geophysics (NRIAG), 11421 Helwan, Egypt
        \and
        Sorbonne Universit{\'e}, CNRS, Institut d’Astrophysique de Paris, 98 bis boulevard Arago, 75014 Paris, France 
        }

   \date{Received 2025 December 18; Accepted 2026 March 20}
  \abstract
   {Properties of the hot intracluster and intragroup medium are mostly set by the underlying gravitational potential well, although complex astrophysical processes at play during their buildup may leave a significant imprint. Observational constraints on the degree and scales of such non-gravitational processes require well-selected samples of objects and deep observations of their gas content.}
   {We aim to study the scaling relation between two global properties of the hot gas, namely its soft-band X-ray luminosity ($L_X$) and its temperature ($T$), by studying a sample of low-mass systems associated with precise redshifts, simultaneously accounting for sample selection biases and associated measurement uncertainties.}
   {This work takes as input a large catalogue of X-ray-selected galaxy clusters (X-CLASS). We perform a thorough revision of the redshifts of sources using deep photometric data from the Legacy Surveys and our own tailored spectroscopic follow-up of 52 low-redshift systems.
   We devise a spectroscopically complete sample of 155 low-redshift ($0.07<z<0.2$) systems, and we measure properties of their X-ray emitting gas, with median $\overline{T}=1.7$~keV and median $\overline{L_X}=10^{43}$~erg\,s$^{-1}$. We infer the relation between $L_X$ and $T$ in a Bayesian framework.}
   {Our sample of groups and clusters with median total mass $\sim 6 \times 10^{13}$~$M_\odot$ reveals a relation $L_X-T$ steeper than predicted by the self-similar model, with a slope $B=3.2 \pm 0.1$. This result fits well within recent studies that together indicate a trend of increasing slope with decreasing median halo mass.}
   {This work supports a scenario of a stronger decrease in luminosity with decreasing mass in the group regime than for massive galaxy clusters. This effect is possibly due to strong and sustained feedback expelling gas efficiently from their relatively shallower potential wells. We release the list of updated redshifts (photometric and spectroscopic) for the full X-CLASS sample and the gas properties of the low-redshift sample. The cluster photometric redshift code presented in the paper {\tt photXclus} is made publicly available\thanks{\url{https://codeberg.org/fsarron/photXclus}}.}


   \maketitle
   \nolinenumbers
   
\section{Introduction}

The intracluster medium (ICM) of galaxy clusters is a hot and fully ionized plasma that emits X-rays through thermal bremsstrahlung. 
X-ray telescopes like XMM-Newton, Chandra, and eROSITA have been built and dedicated to the study of this ICM, which is a significant baryonic component in massive structures like clusters or groups of galaxies \citep{1986Sarazin, 2000Mulchaey}. The distribution of the intracluster gas is recognized as a good tracer of the total gravitational potential and therefore of the total mass $M_{\rm tot}$.   
In the standard self-similar model of cluster formation \citep{1986Kaiser}, the properties of the ICM and the total mass are expected to follow simple scale-free power-law relations derived from gravitational processes. Simple scaling laws like $L_X \propto T^{3/2}$ and $M_{\rm tot}\propto T^{3/2}$ are derived assuming virial equilibrium and bremsstrahlung emission, so X-ray properties like temperature $(T)$ and soft X-ray band luminosity $(L_X)$ are expected to be good mass proxies. However, other physical processes, such as AGN feedback, turbulence, and radiative cooling, are in play in clusters and groups and have a strong impact on the physical properties of the gas and on the mass determination \citep{2012McNamara,2014Nelson}. These non-gravitational heating effects can distort the standard scaling relations by changing the index or the normalization of the power laws. Indeed, X-ray studies have shown steeper relations of $L_X\propto T^{2.7-3.0}$ and $M_{\rm tot} \propto T^{1.6-1.8}$ and a higher value of the intrinsic scatter compared to the self-similar predictions \citep{2015Lovisari, 2016Giles, 2022Bahar, 2024Khalil}. These departures from the original self-similar relations provide compelling evidence of the role of non-gravitational processes in these massive structures, put under the general term `baryonic physics.'

Scaling relations have been extensively studied for massive clusters thanks to various samples observed with recent X-ray satellites \citep[e.g.][]{2009Pratt, 2010Mantz, 2019Bulbul, 2025Ramos-Ceja}. Fewer studies have been conducted on low-mass or low-luminosity galaxy clusters and groups due to their relative faintness \citep{2015Lovisari, 2016Zou, 2022Bahar}. Their study is nevertheless essential to extend our understanding of the physics of structure formation to lower masses. It is expected that the thermodynamic evolution of the ICM is more sensitive to additional baryonic physics like merging processes or AGN feedback for low-mass structures \citep{2021Lovisari} because of their weaker potential well. Moreover, low-mass galaxy clusters are the building blocks of the high-mass galaxy clusters and may have a strong impact on galaxy cluster evolution at cosmological timescales. But selection effects in X-ray or optical observations make it difficult to construct a complete sample down to low masses. Many samples are limited to masses of $\sim 10^{14}$\,M$_\odot$ corresponding to an ICM temperature of 2–3 keV, or $L_X \sim 10^{44}$\,erg s$^{-1}$ typically. The lack of objects with properties below these limits leads to a distortion of the uncertainties distribution, a well-known effect called "up-scattering". All these reasons point towards the importance of mass-selected samples for cosmological studies, although the use of mass proxies has strong limitations. This is also why understanding the dispersion in mass-observable relations provides a wealth of information about cluster physics and cosmological implications, such as the construction of a mass function for structures \citep[e.g.,][]{2012Clerc, 2022Garrel, 2024Ghirardini}.

In order to build samples of galaxy clusters or groups with a large set of physical properties, the redshift is a mandatory quantity in addition to other X-ray or optical observables. Acquisition of cluster redshifts requires extensive optical follow-ups, sometimes very demanding. A first method consists in measuring photometric redshifts provided by multi-band photometric data. This is widely available for many clusters, thanks to imaging surveys covering significant regions of the sky; see, for example, the Dark Energy Survey Instrument (DESI) Legacy Surveys \citep{2019Dey, 2021Zhou}, the Panoramic Survey Telescope And Rapid Response System (PAN-STARRS) survey \citep{2016Chambers}, or the Sloan Digital Sky Survey \citep{2025SDSS-DR19}. Sophisticated cluster search tools are needed, which combine individual redshift estimates to provide a robust value for the whole structure. This method is useful for deriving the redshifts of large samples of galaxy clusters, like the recent results from the extended ROentgen Survey with an Imaging Telescope Array (eROSITA) survey \citep{2024Kluge}. For smaller samples, the analysis can be supplemented with more reliable spectroscopic redshifts.
However, this approach is more limited and difficult to apply to varied and large samples. This remains a major limitation in the use of large samples of galaxy clusters for cosmological purposes. Although, new perspectives are emerging with the arrival of data from the DESI Survey and surveys with new large spectrographs such as the 4-meter Multi-Object Spectroscopic Telescope (4MOST).

In this paper we wish to address the role of low-mass galaxy clusters in samples used for constraining the cluster mass function, in particular the differences between low- and high-mass clusters. To construct our sample, we take advantage of the XMM Cluster Archive Super Survey \citep[X-CLASS,][hereafter: \citetalias{2021Koulouridis}]{2021Koulouridis}, which is based on deep XMM-Newton observations, with rigorous and precise selection criteria.
This paper is organized as follows. In Sect.~\ref{sec:datasample} we describe the parent X-CLASS catalogue as well as our method to obtain updated photometric redshifts from optical data, the spectroscopic follow-up of low-redshift systems, and the assignment of redshifts and confirmation flags to all X-CLASS sources. From there, Sect.~\ref{sec:lowzclusters} describes the selection of a complete sub-sample of low-redshift objects and the measurements of their hot gas properties, whose distribution is modelled (Sect.~\ref{sec:model}) in order to infer the scaling relation between luminosity ($L_X$) and temperature ($T$). The latter constitutes the main result of this study. We summarize our findings in Sect.~\ref{sec:conclusion}.

Throughout this paper, $L_X$ stands for the soft-band X-ray luminosity calculated in the $0.5-2.0$ keV energy band within an aperture $R_{500}$ around the cluster center\footnote{A radius $R_{500}$ encloses an amount of matter whose average density is 500 times the critical density of the Universe at the cluster redshift.}. We use the following cosmological parameters: $\Omega_{\rm m} = 0.3, \ \Omega_\Lambda = 0.7, \ H_0 = 70 \ \mathrm{km \ s}^{-1} \ \mathrm{Mpc}^{-1}$. Brackets $\langle \cdot \rangle$ denote an average value, and overlines $\overline{\cdot}$ a median value.

\section{Datasets, parent catalogue and redshift upgrade\label{sec:datasample}}

    \subsection{The X-CLASS parent sample\label{sec:xclass}}

X-CLASS is a search for X-ray-emitting galaxy clusters in the imaging data of the European high-energy mission XMM-Newton. The survey takes advantage of the large field of view of the EPIC cameras (equipped with MOS1, MOS2, and PN detectors) associated with the large collecting area of the three telescopes and with the $\sim 15\arcsec$ point-spread function (PSF) at keV energies.
Putting together about $270 \deg^2$ of archival pointings, \citetalias{2021Koulouridis} published a sample of 1646 galaxy cluster candidates discovered in XMM data acquired during the time frame 2000--2015.

We briefly recall the steps involved in the construction of the cluster catalogue and redirect to \citetalias{2021Koulouridis} for details. Images in the 0.5--2\,keV band are first filtered with wavelets, and a list of putative sources and their positions on sky is obtained by running the \texttt{SExtractor} software \citep{1996Bertin}. An algorithm then fits a range of source models to raw X-ray cutouts extracted around each source candidate location. Model significance values derive from the standard "C-statistic" \citep{1979Cash}. Each entry is associated with a detection likelihood, representing the C-stat excess of a PSF model relative to a background-only model. The extent likelihood quantifies the C-stat excess of an extended source relative to a PSF model. In this context (i.e., detection), an extended source model consists of a $\beta$-model \citep{1976Cavaliere} with core radius (or extent) left as a free parameter; source characterization (Sect.~\ref{sec:measurements Lx Mgas}) will bring refinements to this surface brightness model. After removal of duplicates -- due e.g.~to overlapping pointings or split detections --, sources with values of extent, extent likelihood, and detection likelihood exceeding predefined threshold values are listed in a primary catalogue.
The sample selection function follows from Monte Carlo realizations of mock XMM pointings incorporating realistic X-ray backgrounds and Active Galactic Nuclei populations as well as various galaxy cluster shapes injected at random \citep[][]{2006Pacaud, 2012Clerc}. Mock images are processed identically to real pointings. A selection model is then constructed such as to output the expected fraction of objects of a given size and flux passing the aforementioned selection cuts.

A final visual inspection of the X-ray images (aided with shallow optical imaging) enabled to establishment of the list of sources that make up the X-CLASS catalogue released by \citetalias{2021Koulouridis} and published in an open database\footnote{\url{https://xmm-xclass.in2p3.fr/}}. Cross-matching with literature data, in particular from the NASA Extragalactic Database (NED) and our own previous follow-up campaigns \citep{2017Ridl, 2020Clerc}, enabled us to obtain the redshift of 982 objects, as presented in \citetalias{2021Koulouridis}. 
All sources were thus classified with regard to their redshift estimate with a status selected among:
\begin{itemize}
    \item Confirmed: three or more galaxies with concordant redshift within the $R_{500}$ radius; or the redshift of the BCG (Brightest Cluster Galaxy) is confirmed.
    \item Tentative: One or two galaxies with concordant redshift and BCG can not be securely identified 
    \item Photometric: Only photometric information available, from the literature or from our own observations \citep{2017Ridl} 
    \item Provisional: Very tentative identification of a cluster 
    \item No redshift: all remaining cases 
\end{itemize}
Prior to the present study, only 60\%\ of the X-CLASS clusters were spectroscopically confirmed.
Next we describe our efforts in increasing the redshift completeness of the sample together with the precision of photometric redshifts.

\subsection{Photometric redshifts of X-CLASS clusters\label{sec:photoz}}

\subsubsection{Photometric data}

To increase the redshift completeness of the X-CLASS database, we used the DESI Legacy Survey\footnote{\url{https://www.legacysurvey.org/}} (LS) Data Releases DR9 and DR10 \citep{2019Dey}. This wide large-band photometric survey covers more than 14,000 deg$^2$ both in the Northern and Southern sky, in up to 4 optical bands ($g, r, i, z$). Observed sources are matched with the unWISE all-sky catalogue \citep{2021Zhou} to obtain the photometry in 3.4 (W1) and 4.6 microns (W2) bands, such that multi-color catalogues covering optical and infrared can be built.
Sources are detected using a matched filter on the stacked images \citep{2012Lang}, and multi-band photometry is measured on individual exposures.  
Photometric redshifts for individual sources are then computed using a Random Forest algorithm described in \citet{2023Zhou}.

At $z_{\rm mag} < 21$, the Normalized Median Absolute Deviation (NMAD) 1$\sigma$ photo-$z$ uncertainty for the DR10 (DR9) is $\sigma_{\rm NMAD} = 0.017~(0.022) \times (1 + z_{\rm spec})$ when compared to the DESI Science Verification (SV) Luminous Red Galaxies (LRG at $0.3 \lesssim z \lesssim 1$) sample \citep{2023Zhou}; and $\sigma_{\rm NMAD} = 0.015~(0.019) \times (1 + z_{\rm spec})$ when compared to the Bright Galaxy Survey (BGS at $0.05 \lesssim z \lesssim 0.5$). DR10 provides additional $i$-band photometry, which improves the quality of the photometric redshifts\footnote{\url{https://www.legacysurvey.org/files/dr10_photoz_performance.txt}}. Taking $|{\Delta z}| = |z_{\rm phot} - z_{\rm spec}|$, for the same samples, the outlier fractions for $\eta (|{\Delta z}| > 0.10)$ are 1.1\% (1.5\%) and 0.9\% (1.3\%), respectively.

\subsubsection{Cluster redshift estimate}
The X-CLASS clusters photometric redshifts were estimated using a combination of two photometric redshift codes: {\tt photXclus}\footnote{\url{https://codeberg.org/fsarron/photXclus}}, a Bayesian cluster redshift code we developed for the purpose of the present study, and the {\tt RedMaPPer} cluster finder used in scan mode. Both methods are presented extensively in Appendix~\ref{sec:photoz-annex}.

{\tt photXclus} uses photometric redshift estimates of individual galaxies, determined by \citet{2021Zhou} for all galaxies in the LS DR9. For X-CLASS clusters falling in the LS DR9 footprint, photometric redshifts of the clusters are estimated using a Bayesian analysis based on individual galaxy photo-$z$ in the area around the X-ray sources. 

{\tt RedMaPPer} (red-sequence Matched-filter Probabilistic Percolation) is an optical cluster finder algorithm \citep{2014Rykoff}. It was extended and enhanced to run on the LS DR9 and DR10 \citep{2020IderChitham, 2024Kluge}. {\tt RedMaPPer} in scan mode estimates the redshift, richness, and central position of each cluster using a combination of three filters that model (i) the radial profile, (ii) the luminosity distribution, and (iii) the color distribution of galaxies. These filters are applied in the cluster vicinity, both on red-sequence cluster galaxies and on background galaxies. The redshift is estimated through the redshift dependence of the luminosity and color models for cluster galaxies.

\begin{figure*}
    \centering
    \includegraphics[width=0.9\linewidth]{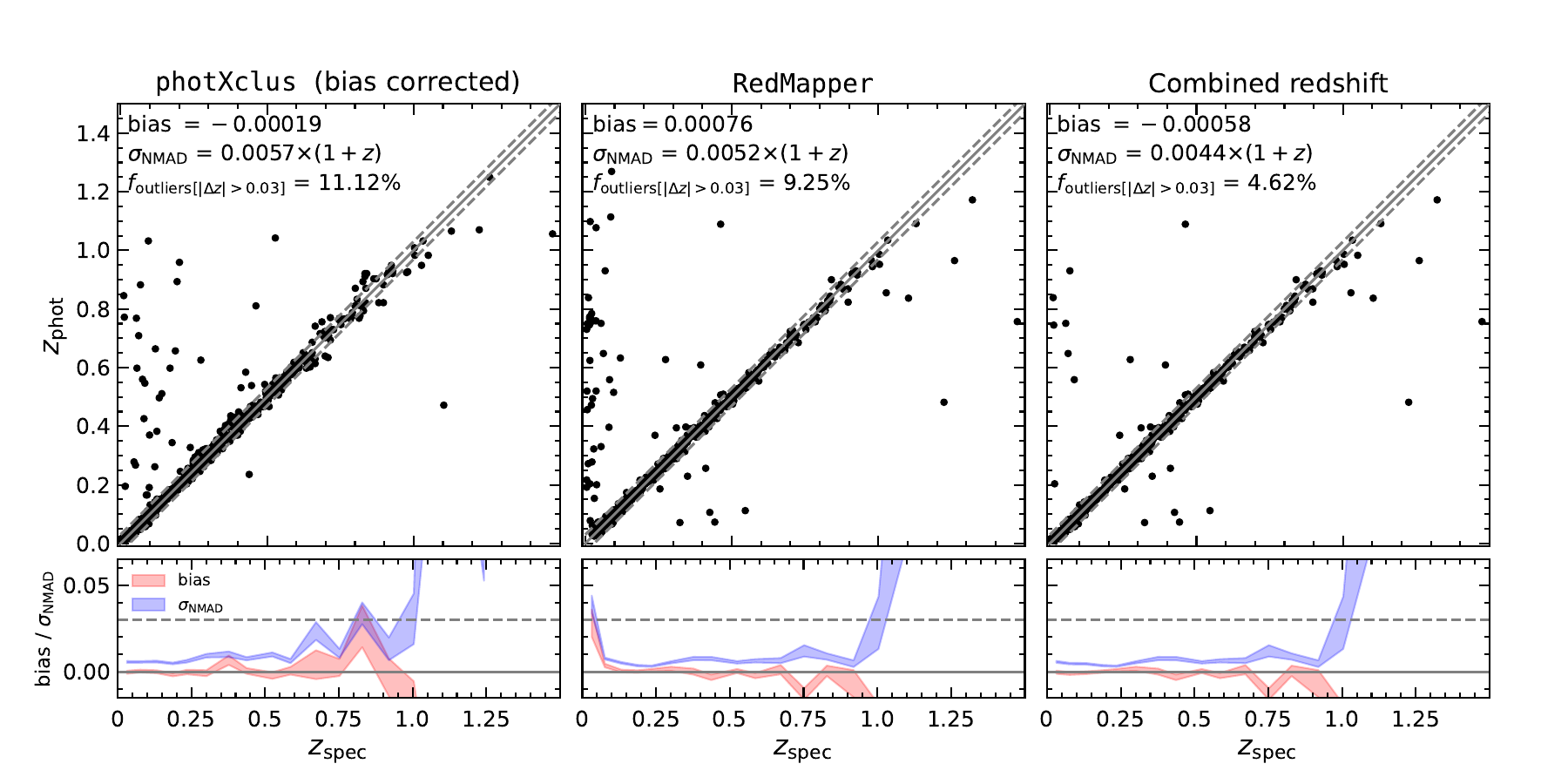}
    \caption{Performances of the photometric redshift estimates. {\it Top}: photometric redshifts vs. spectroscopic redshifts for confirmed X-CLASS clusters. {\it Bottom}: Bias (red) and NMAD scatter (blue) of the photometric redshift estimates as a function of spectroscopic redshifts for confirmed X-CLASS clusters. From left to right: (i) bias-corrected {\tt photXclus} redshift, (ii) {\tt redMaPPer} redshift, and (iii) best photometric redshift estimate. The quantities $\sigma_{\rm NMAD}$, bias (defined as the median of $\Delta z = z_{\rm phot} - z_{\rm spec}$), and the outlier fraction are defined in the main text.}
    \label{fig:photo_z}
\end{figure*}

To obtain the final redshift estimate, we combine the outputs of {\tt photXclus} and {\tt RedMaPPer} in order to minimize bias, scatter, and the outlier fraction when compared to X-CLASS clusters with spectroscopic redshifts (Fig.~\ref{fig:photo_z}). Overall, {\tt RedMaPPer} has better performance, except at low redshift (z < 0.15), where the outlier fraction is high, leading to higher bias and scatter. Conversely, while {\tt photXclus} scatter and outlier fractions are slightly larger overall, {\tt photXclus} avoids the outlier population at low redshift seen in {\tt RedMaPPer}. This is because {\tt photXclus} uses the extent of the X-ray signal to inform the cluster redshift prior (see Appendix~\ref{sec:photXclus} for details). Therefore, for our final combined photometric redshift estimate, if $z_{\rm {\tt photXclus}} < 0.15$, we use the {\tt photXclus} photo-$z$; otherwise, we use the {\tt RedMaPPer} photo-$z$. 

The number and fraction of clusters with an estimated photometric redshift for each code and the combined photometric redshift estimate are presented in Table~\ref{table:redshifts_summary}.
Our combined photometric redshift estimate has an NMAD scatter $\sigma_{\rm NMAD}$ of $0.0044 \times (1 + z)$, a bias (defined as the median of $\Delta z = z_{\rm phot} - z_{\rm spec}$) of $-5.8 \times 10^{-4}$, and an outlier fraction $f_{\rm outliers}(|\Delta z| > 0.03) = 4.62\%$ when compared to X-CLASS clusters with spectroscopic redshifts, an improvement over using either of the two photo-$z$ codes individually (see Fig.~\ref{fig:photo_z}).

\begin{table*}
\caption{Number of clusters with a photometric redshift estimated by each photo-$z$ code, and the combined catalog.}
\label{table:redshifts_summary}
\centering 
\begin{tabular}{cc|ccc|ccc|ccc}
\hline
\hline
\multicolumn{2}{c}{} & \multicolumn{3}{c}{\tt photXclus} & \multicolumn{3}{c}{\tt redMaPPer} & \multicolumn{3}{c}{combined} \\
\smallskip
\citetalias{2021Koulouridis} status flag & $N_{\rm X-CLASS}$ & $N_{\rm DR9}$ & $N_{{\rm photo-}z}$ & $f_{{\rm photo-}z}$ & $N_{\rm DR10}$ & $N_{{\rm photo-}z}$ & $f_{{\rm photo-}z}$ & $N_{\rm DR10}$ & $N_{{\rm photo-}z}$ & $f_{{\rm photo-}z}$ \\
\hline
\textit{Tentative}   & 94  & 80  & 75 & 80\% & 91  & 85 & 90\% & 91  & 90 & 96\% \\
\textit{Photometric} & 202 & 134 & 132 & 65\% & 184 & 173 & 86\% & 184 & 176 & 87\% \\
\textit{No redshift} & 281 & 197 & 175 & 62\% & 213 & 209 & 74\% & 260 & 236 & 84\% \\
\hline
\end{tabular}
\tablefoot{Only clusters with no confirmed redshift are considered. The first column refers to the confirmation status from \citetalias{2021Koulouridis}, {\it i.e.} prior to this work.}
\end{table*}

\subsection{Spectroscopic survey with MISTRAL\label{sec:spectroz}}

\subsubsection{Observations}
To complete the sample of spectroscopically confirmed clusters at low redshift, we set up a dedicated observing program using the T193 facility at the Observatoire de Haute-Provence. A total of 22 nights was allocated to the program with the MISTRAL instrument \citep[Multi-purpose InSTRument for Astronomy at Low-resolution][]{2024Schmitt}. We used the Blue mode, which offers a useful spectral range of 4500 -- 8000 \AA\ and a spectral resolution of $\sim 700$. The slit width is fixed at 1.9\arcsec, a value suited to the site's average seeing conditions. The survey was split into seven runs between October 2021 and February 2025 during dark time. We estimate that, over the 22 nights of observation, 2/3 of the time was spent under acceptable observing conditions (clear skies and seeing below 4\arcsec). For each pointing, the exposure time was limited to 30 minutes at most to ensure secure telescope guiding. If needed for faint objects, observations were duplicated to achieve total integration times of 1–2 hours, depending on the magnitudes of the galaxies.

The targets were chosen from the X-CLASS sources as low redshift candidates with a status No redshift, Photometric, or Tentative, in order of priority (as defined in Sect.~\ref{sec:xclass}). During the first runs, the new photometric redshifts were not available, and we relied mainly on the available images in the LS DR9. Additional clusters with a declination between $-7 \deg $ and $0 \deg $ were added as fillers when required. Note that the high latitude of Observatoire de Haute Provence provides access to sources with the highest declination, which were also considered priority targets. In total, more than 55 different X-CLASS sources were observed during the program.

\subsubsection{Data reduction and redshift measurements}
We developed a semi-automated, end-to-end pipeline to reduce and extract the spectra. We used the PyRAF environment, which is a Python-based command language for IRAF\footnote{\url{https://iraf-community.github.io/pyraf.html}}, as well as standard procedures for long-slit spectroscopy.
Wavelength calibration was obtained with an accuracy of less than 0.8 \AA\ across the entire 2D image, and spectra were re-sampled with 2 \AA\ bins from 4500 \AA\ to 8000 \AA. Sky subtraction was applied to the entire 2D image prior to source extraction. We proceeded with manually identifying and extracting the selected sources within the slit.
Finally, we carried out flux calibration, but it was subsequently little used due to poor flux restitution below 5000 \AA\ or above 7500 \AA. After carefully identifying the objects in the slits, we co-added the spectra to form the final spectrum of each object. In total, we extracted more than 200 independent objects, including stars.

Nearly all the galaxies observed in the program are early-type, passive galaxies that were selected to be the brightest members of our groups and clusters, and only 10\% of the spectra show emission lines. Therefore, the spectra are dominated by absorption line features, which are more difficult to detect in spectra with a poor signal-to-noise ratio. We used a combined approach to identify true spectral features by excluding sky residuals and other instrumental defects and then measured the redshift. First, we used the cross-correlation module in PYRAF ({\tt noao.rv.fxcor}), which computes a Fourier cross-correlation, as defined by \citet{1979Tonry}.
A solution was provided in 85\%\ of cases where the identification was not a star and the spectrum did not contain emission lines. We used the elliptical-type template distributed by the CfA\footnote{\url{http://tdc-www.harvard.edu/iraf/rvsao/templates/}} for these galaxies. In parallel, we also used the {\tt Marz} software\footnote{\url{https://samreay.github.io/Marz/}} \citep{2016Hinton}, which is available through an online graphical interface. We used a simple approach, visually confirming the fit between the "true" spectral features in our spectra and those in the "Early type Absorption galaxy" template available in the interface, after applying the smallest level of smoothing to reduce the noise. We checked that both redshift measures were coincident. In the general case when the difference between both methods was smaller than 0.0005, we fixed the redshift to the value determined by {\tt fxcor} and used the value measured with {\tt Marz} otherwise. In the very few cases where individual exposures were not too noisy, we also checked that the correlation tool was giving concordant redshift values within the same interval of 0.0005. All in all, the global accuracy of redshift measurements is $\delta z \sim 0.001$, which corresponds to the expected uncertainty with this low-resolution spectrograph.

\begin{figure}
    \centering
     \includegraphics[width=\linewidth]{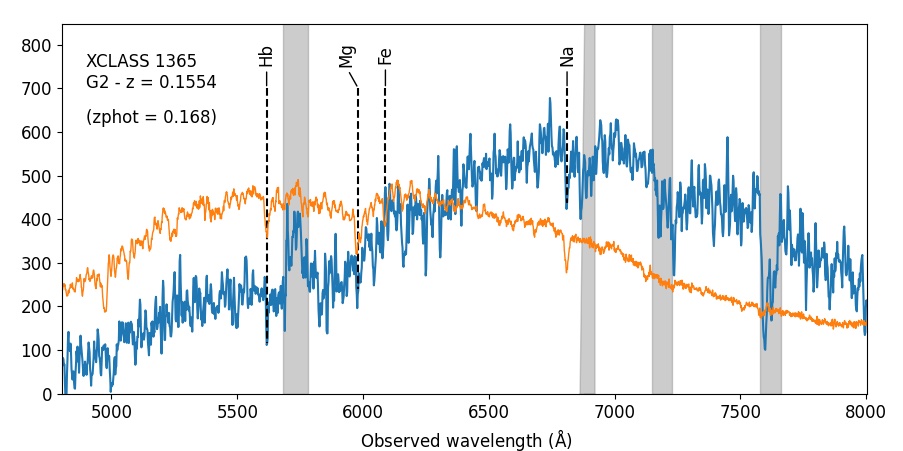}
      \includegraphics[width=\linewidth]{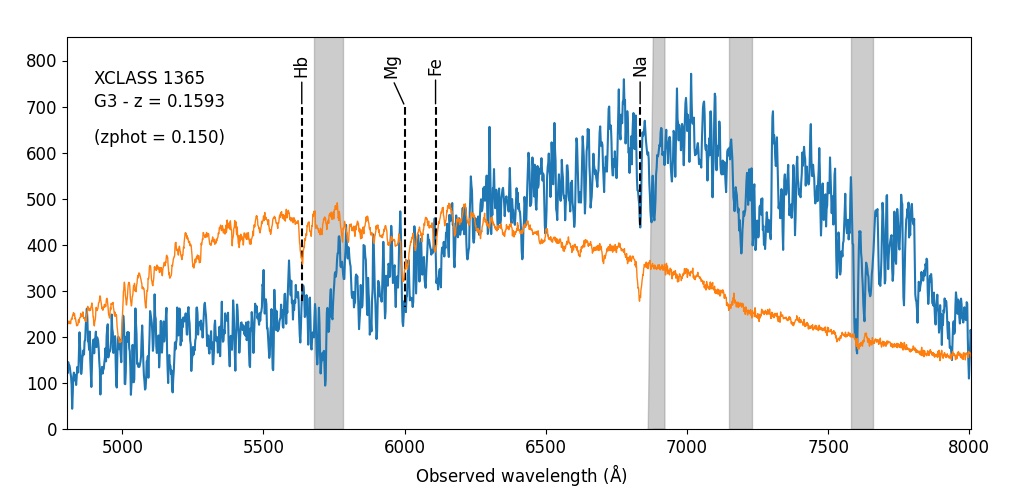}
    \caption{Spectra of two galaxies (G2 and G3) in xclass1365 as observed with MISTRAL. The redshift measurements are based on the main absorption lines detected and identified using the template spectrum of an early-type galaxy (in orange, redshifted to $z=0.1554$ for G2 and $z=0.1593$ for G3 as measured with the cross-correlation tool {\tt fxcor}). The intensities are in arbitrary units, and neither spectrum is flux calibrated. The shaded areas correspond to atmospheric absorption lines except around 5700 \AA, where there is a defect in the MISTRAL optics. Both spectra have been graded with a confidence level of 3 for the redshift measurement.}
    \label{fig:sp_mistral}
\end{figure}

The final list of galaxies for which a redshift was measured in the MISTRAL program contains 144 objects. Stars have been removed from this list presented in Appendix \ref{sec:zmistral}. Typical examples of observed spectra are shown in Figure~\ref{fig:sp_mistral}. Note that except for one galaxy identified as a radio source (3C300, $z=0.273$) in the xclass21544 cluster, no quasar has been measured spectroscopically with MISTRAL. All BCGs are clearly elliptical galaxies, so we assert that among the low redshift XCLASS clusters, contamination by active galaxies is at its lowest. Finally, there is good agreement with the photometric redshifts from the DESI LS, with an RMS difference of 0.033, or an NMAD uncertainty $\sigma_{\rm NMAD} = 0.015$ for the whole MISTRAL sample, a value fully compatible with the general one of the DR9 catalog (Section \ref{sec:photoz}).

\subsubsection{Galaxies identification and cluster/group redshift validation}
\begin{figure}
    \centering
    \includegraphics[width=\linewidth]{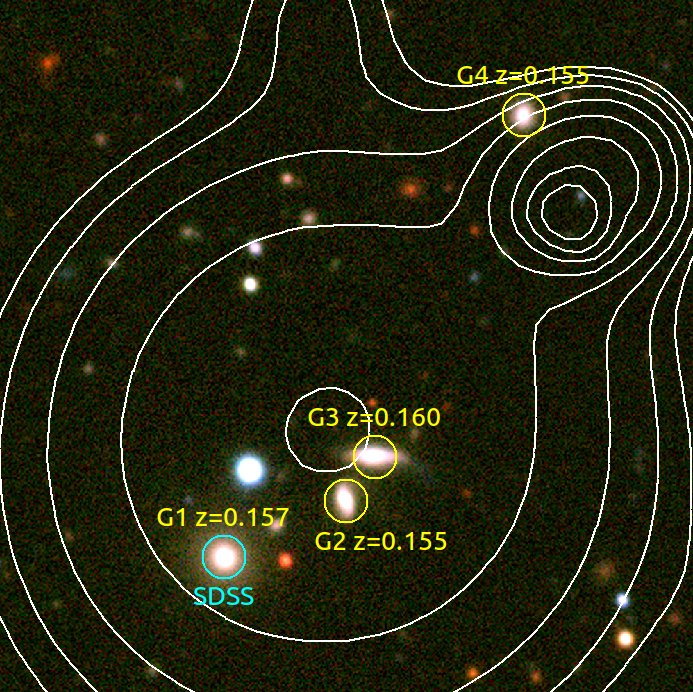}
    \caption{Three-color image of xclass1365 extracted from the LS DR9. The yellow circles represent galaxies observed with MISTRAL, along with their redshifts. The blue circle indicates the source was also measured in the SDSS (with concordant redshift). North is up and east is left, and the field of view is 1.87\arcmin\ wide. The white contours correspond to the X-ray contours smoothed and filtered with a wavelet algorithm.}
    \label{fig:xclass3135}
\end{figure}
Galaxies were identified in the LS DR9, for which coordinates and magnitudes are available for each detected object. When a cluster falls outside the LS footprint, images were taken from the Pan-STARRS1 imaging survey available through the ESA Sky environment\footnote{\url{https://sky.esa.int/esasky/}}. Additional photometric redshift information was retrieved for the galaxies observed with the LS DR9. This information was used to assist with identification and is included in the full table (Appendix \ref{sec:zmistral}). Figure~\ref{fig:xclass3135} shows an example of the final identification.

A total of 52 different clusters were identified in the MISTRAL program, with a redshift determined for each one. We removed all objects from the list that were observed with no conclusive identification. Additional spectroscopic data from the 6dF or SDSS surveys \citep{2009Jones, 2025SDSS-DR19} were available in a few cases and were added to the sample to improve cluster identification and redshift validation.
The cluster redshift was assigned, including the flag "S" described in Appendix \ref{sec:redshiftable}. If the BCG has a measured redshift and the other galaxies have common values, then $z_{cl} = z_{BCG}$. However, if the BCG is not well identified or has no measured redshift, then $z_{cl} = \langle z \rangle$ from all available redshifts. A summary of the redshift assignments to the X-CLASS clusters is provided in Table \ref{table:cluster_redshifts}. 

\subsection{Updating redshifts in the X-CLASS database \label{sec:redshifts}}

To improve the quality and completeness of the X-CLASS database, we first included all the new redshifts available from the MISTRAL runs. We then updated the final redshift for all sources, aggregating the various available information such as the redshifts available in the field of each cluster, the photometric redshift as determined in Sect.~\ref{sec:photoz}, and the concordance between values. We thus propose a new redshift value, accompanied by an additional column containing a three-digit flag named "ZSP", where each digit is defined by different criteria. This flag is defined so that it can be converted to a human-readable status that is similar to the status defined by \citetalias{2021Koulouridis}. A complete description of the procedure is provided in Appendix~\ref{sec:redshiftable}.

In summary, 1599 X-CLASS clusters are left in the new catalogue provided in Table~\ref{table:fullredshifts}, with a final status corresponding to Confirmed or Photometric for most of them. Only a very few cases of Tentative classification remain (9 sources), and we still have 70 clusters with No redshift. We list below the following correspondences between the initial status, our "ZSP" classification, and the number of clusters in each category: 
\begin{itemize}
    \item Confirmed (1060 clusters) : flags 12x, 14x, 15x, 22x, 23x, 24x, 25x 
    \item Tentative (9 clusters) : flags 110, 130
    \item Photometric (460 clusters) : flags 3xx, 41x 
   \item No redshift (70 clusters) : flag 000
\end{itemize}

\section{Sample selection and physical properties of low redshift X-CLASS clusters}
\label{sec:lowzclusters}
\subsection{The X-CLASS low redshift sample (X-CLASS-LR)\label{sec:sample}}

\begin{figure*}
    \centering
    \includegraphics[width=0.45\linewidth]{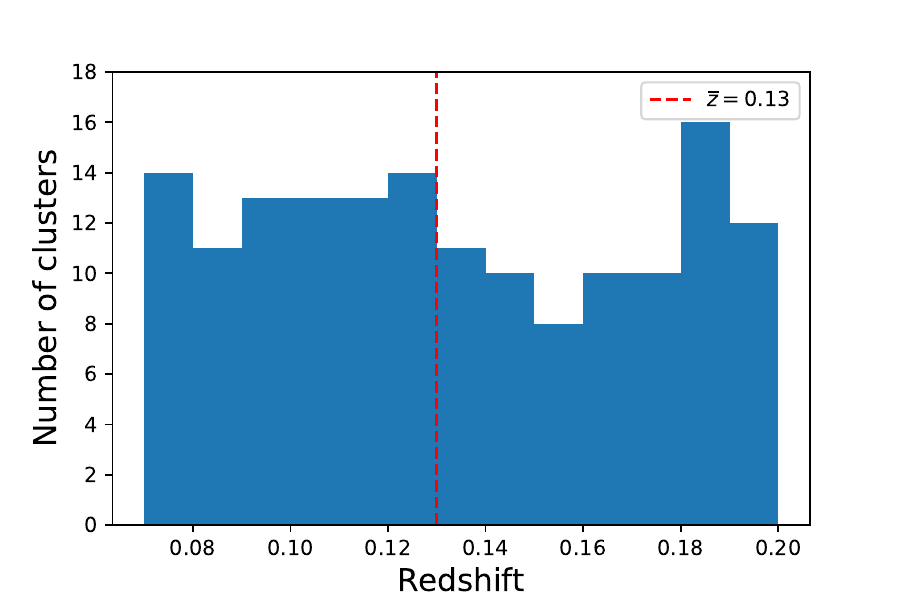}
    \includegraphics[width=0.45\linewidth]{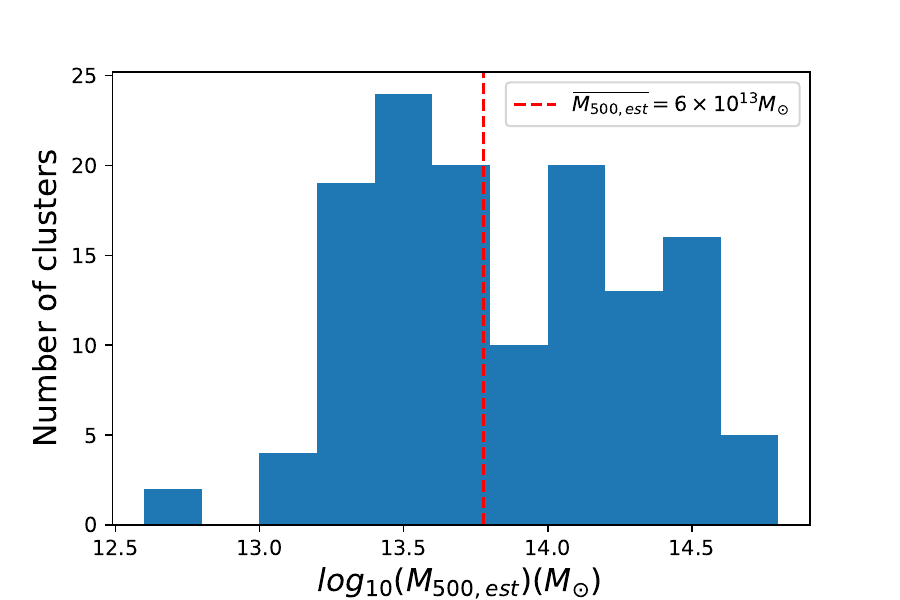}
    \caption{Left: Redshift distribution of the X-CLASS-LR sample composed of 155 groups and clusters of galaxies. The dashed line indicates the median redshift. Right: Estimated mass distribution of the X-CLASS-LR sample. The dashed lines indicate the median redshift and the median mass (logarithmic scale).}
    \label{fig:zdist}
\end{figure*}

Equipped with results developed in the previous section, we are in a position to clearly define a low-redshift sub-sample of clusters extracted from the X-CLASS database that meets the following criteria: declination $0\deg <\delta<+90\deg$, a "ZSP" flag that is not equal to 000, and a new redshift in the range $0.07\leq z\leq0.2$. This sample was defined that way, so we included all the new redshifts obtained in the MISTRAL program; therefore, only clusters in the northern hemisphere are selected. Moreover, at very low redshift, most sources corresponding to extended X-ray sources are either nearby galaxies or groups with an extent much larger than the XMM-Newton field of view. X-CLASS is therefore not optimized to study these targets because of unclear selection biases. This is why we discarded all sources at redshift smaller than 0.07. 
Application of these criteria leads to a sample of 157 clusters, with only two of them with uncertain photometric redshift (ZSP = 302) that have been discarded. 
Ultimately, we are left with a sample of 155 clusters that has a complete spectroscopic identification (Figure \ref{fig:zdist}). In what follows, we call this sample "X-CLASS-LR".

\begin{figure}
    \centering
    \includegraphics[width=1.0\linewidth]{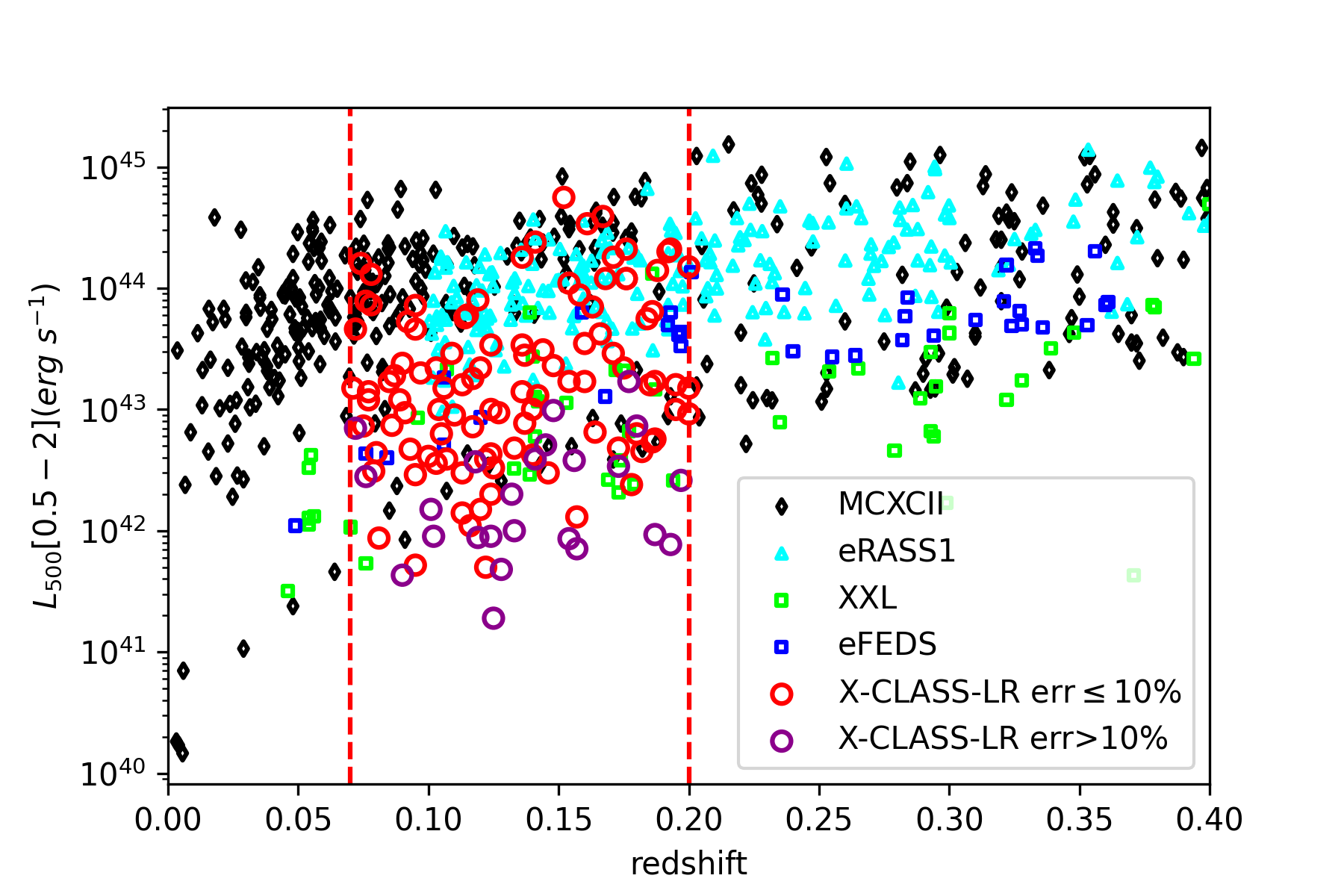}
    \caption{X-ray luminosity ($L_X$) versus redshift for the X-CLASS-LR sample (red circles). X-CLASS-LR sources with measurement uncertainties greater than 10\%\ are plotted in purple. Red dashed lines represent the redshift limits of the sample. Additional samples with measurement uncertainties below 10\% are shown: the MCXC-II survey in black \citep{2024Sadibekova}, the Cosmology Sample of the eROSITA DR1 in cyan \citep{2024Bulbul}, the XXL DR2 survey in green \citep{2018bAdami}, and the eROSITA Final Equatorial-Depth Survey (eFEDS) sample in blue \citep{2022Liu}.}
    \label{fig:Lx-z}
\end{figure}

The X-CLASS-LR sample is dominated by low-mass and low-luminosity clusters and groups. Figure \ref{fig:Lx-z} presents the X-ray luminosity distribution $L_X$ with respect to the redshift. The next section presents the detailed processing of X-ray data that results in the measurement of these luminosities, as well as other physical parameters. It should be noted that, of the 155 sources, only 24 clusters have an error on the $L_X$ measurement larger than 10\%. Additional fitted parameters may also be subject to larger errors, which will be taken into account in the global analysis of the sample. 
Interestingly, our sample compares well with the eROSITA Final Equatorial-Depth Survey (eFEDS) sample, which was built using data acquired during the performance and verification phase of the eROSITA mission \citep{2022Liu}. 
The area and the mass range are similar within the considered redshift range. However, XCLASS observations are 3 to 6 times deeper, and the smaller PSF of XMM-Newton leads to the detection of more low-mass clusters.

\subsection{X-ray data spectral analysis \label{sec:properties}}
\begin{figure}
    \centering
    \includegraphics[width=\linewidth]{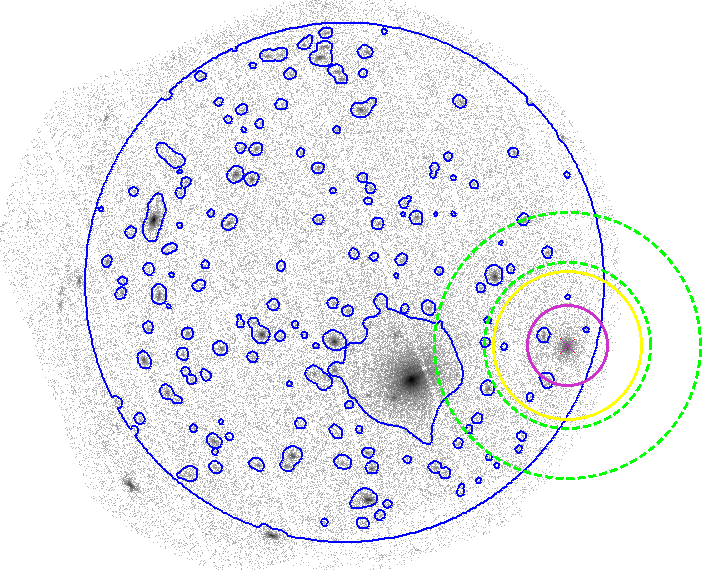}
    \caption{$30\arcmin \times 30\arcmin$ image ($0.5-2$~keV) of the XMM pointing hosting the xclass1769 detection. The red cross shows the X-ray center of the cluster. The magenta circle shows the initial radius $R_{\rm fit}$ used for the spectral fit. The yellow circle shows the estimated radius $R_{500}$ $=450~kpc = 223 "$. The green dashed annulus marks the background extraction region, and blue contours indicate masked sources and regions outside the considered field of view.}
    \label{fig:observation}
\end{figure}

Gas temperature and metallicity for each cluster were measured with the Science Analysis Software (SAS)\footnote{\url{https://www.cosmos.esa.int/web/xmm-newton/sas}} of XMM-Newton. The X-ray spectrum is extracted inside a given cluster radius, while the background is extracted in an annulus whose size is scaled to the cluster radius, after masking the emission from other X-ray sources (Fig.~\ref{fig:observation}). However, temperature and metallicity are not expected to be homogeneous within the whole cluster, so their measured values will depend on the extraction radius. We wish to use $T_{500}$ and $Z_{500}$, measured within the radius $R_{500}$. This radius has to be determined first, through an iterative procedure based on \citet{2020Molham}: the initial radius $R_1$ is set to $R_{\rm fit}$, defined as the radius where the X-ray surface brightness of the cluster is indistinguishable from background fluctuations. Temperature and metallicity are then fitted from spectra extracted within $R_1$. A new radius $R_2$ is computed from the $T_{500}-R_{500}$ scaling relation of \citet{2015Lovisari} for their total sample. If $|R_1-R_2|\leq 1 \arcsec$, a threshold chosen as a reasonable tolerance, the iteration stops and the corresponding value is adopted. Otherwise, the procedure is repeated with $R_2$ as the updated radius until final convergence. The final value is taken as a rough estimate of $R_{500}$, for which we do not compute uncertainties given the series of approximations it involves.

X-ray spectra are fitted with {\tt XSpec} (version 12.13.1) \citep{1996Arnaud}, taking into account Galactic absorption with the neutral hydrogen column density $N_H$ \citep{2016H14PI} and the redshift $z$ of the cluster. $T_{500}$ and $Z_{500}$ are fitted simultaneously, and their uncertainties are computed. The abundance of heavy elements is quoted relative to the solar abundance table of \citet{2000Wilms}.

Although the signal-to-noise (S/N) ratio of the X-ray data is low in most cases, the measure of the metallicity can prevent the so-called "Fe bias", which is a degeneracy between temperature and metallicity in spectral fits of galaxy clusters hosting multi-phase gas. This is particularly sensitive for low-mass groups and clusters because at low temperature, X-ray emission is dominated by metallic emission lines compared to the Bremstrahlung continuum emission \citep{1998Buote, 2000Buote, 2021Gastaldello, 2022Mernier}. If the metallicity was fixed to a given standard value, spectral fit would overestimate the temperature in the case of low metallicity and would introduce a significant, undesired bias in the temperature measurement. Only in the few cases where the spectrum S/N ratio is too low we fix the metallicity to the value $Z_{500}=0.3\,Z_{\odot}$ and fit the temperature $T_{500}$ only. The values of the temperature $T_{500}$, the metallicity, the radius $R_{500}$, and their associated errors are reported in Table \ref{table:measurements}. The measured temperatures range from 0.5 to 6 keV with a median value $\overline{T}=1.7$\,keV, and the measured metallicity has a median value $\overline{Z}=0.26\,Z_{\odot}$, considering only properly converged spectral fits.
In 64\% of the measured clusters, the relative uncertainty on $T_{500}$ is lower than $10\%$.

The distribution of the X-ray temperatures shows that 57\% of clusters in the X-CLASS-LR sample have $T_{500}< 2$ keV, a value that is characteristic of low-mass clusters. 
From these temperature measurements we can compute a mass estimate $M_{500, {\rm est}}$ (Fig.\ref{fig:zdist}) using the $T_{500}-M_{500}$ relation of \citet{2015Lovisari} (their full sample). Our cluster mass estimates spread between $0.5$ and $55\times 10^{13}$\,M$_{\odot}$ with a median value of $\overline{M_{500}}=6\times 10^{13}$\,M$_{\odot}$. About $60\%$ of them have an estimated mass below $10^{14}$\,M$_{\odot}$ and therefore lie in the galaxy group regime.

\subsection{X-ray luminosity}
\label{sec:measurements Lx Mgas}

The X-ray luminosity of the X-CLASS-LR sample is measured after careful inspection of the XMM-Newton images.
De-projection of the 2-dimensional (2D) surface brightness distribution is needed to get the 3D electron density distribution $n_e (r)$ from which $L_X$ is deduced.
We forward-fitted the observed 2D surface brightness assuming a 3D electron density model with a Markov Chain Monte Carlo algorithm. {\tt MBProj2D} \citep{2018Sanders} is a Python package developed for this purpose, taking advantage of multiple observations in several bands. We used the code in the single-band mode, namely using the energy range [0.5--2] keV that optimizes the S/N ratio of clusters and groups observed with XMM-Newton. 
The advantage of {\tt MBProj2D} is that it models clusters in 3D, making the clusters inhomogeneities easier to take into account compared to a 1D model. It makes easy the implementation of PSF flux redistribution, and it includes a parameterization of the background level. The data are fitted assuming a gas temperature $T=T_{500}$ as previously determined. In agreement with \citet{2009Sun}, the metallicity profile of our model for groups is dominated by slightly sub-solar abundances, which we take equal to $Z=0.3\,Z_{\odot}$, with a sharp $Z \simeq Z_{\odot}$ peak at the center. Because our data involves faint sources, we fixed the 3D electron density profile $n_e (r)$ to a simple $\beta$-model. This assumption reduces the number of parameters to fit to the following three physical parameters: the central electron density $n_{e,0}$, the core radius $r_c$, and the slope $\beta$. The images coming from the three XMM-Newton detectors are fitted simultaneously, letting each background value be a free parameter ($b_{MOS1}$, $b_{MOS2}$, and $b_{PN}$). Convolution by the XMM-Newton PSF is also included in the model. {\tt MBProj2D} then computes the projected surface brightness image that is compared to the observed one with a Poisson likelihood. The MCMC sampler provides an output that includes the uncertainties of the fit. We note that degeneracies occur between the estimates of $\beta$ and $r_c$. It is a well-known result of fitting a $\beta$-model to low-surface-brightness objects \citep{1999Mohr}. Within the full sample of clusters for which the fit converged, we find the median values $\overline{\beta}=0.58$, $\overline{r_c}=64$\,kpc, and $\overline{r_c/R_{500}}=0.1$.
This is consistent with the few previous studies using $\beta$-models \citep[e.g.,][]{2010Alshino}.

Finally, the total X-ray luminosity in the [0.5-2] keV band $L_X$ is computed by integrating the inferred emissivity profile up to the radius $R_{500}$. This step takes into account the value of $T_{500}$ for determining the conversion factor between photon counts on the XMM-Newton detectors and physical fluxes, as well as the redshift. The propagation of uncertainties on the free parameters of the model is also performed.
The fit of the profile could not converge in only eight clusters because of the low signal-to-noise ratio of the X-ray signal in the images, so we chose to fix the slope to its standard value $\beta={2}/{3}$ for these clusters to get the estimate of $L_X$. The measured luminosities range from $1.9\times10^{41}$ to $5.6\times 10^{44}$~erg\,s$^{-1}$ with a median value $\overline{L_X}=1\times10^{43}$~erg\,s$^{-1}$. It is quite similar to the low-mass part of the eFEDS sample \citep{2022Liu} that ranges from $3\times 10^{41}$ to $4\times 10^{44}$~erg\,s$^{-1}$ with a median value $\overline{L_X}=3\times10^{43}$~erg\,s$^{-1}$. These values will be used for comparison in the rest of the paper. In 80\% of the clusters, the relative uncertainties on $L_X$ are lower than 10\%. The measured X-ray luminosity and their associated errors are reported in Table~\ref{table:measurements} as part of App.~\ref{sec:measurements}. 

\section{Modeling the $L_X - T$ relation in the X-CLASS-LR}
\label{sec:model}
\subsection{Pre-selection of the working sample}

\begin{table}[H]
\caption{Various selection steps leading up to the constitution of the final sample used in the model.}
\begin{tabular}{ll}
\hline\hline\noalign{\smallskip}
Sample & N$_{clusters}$ \\
\noalign{\smallskip}\hline\noalign{\smallskip}
X-CLASS Full sample \citepalias{2021Koulouridis} & 1646 \\
 $\hookrightarrow$ X-CLASS with at least photo-z (this work) & 1523 \\
 \: $\hookrightarrow$ {\bf X-CLASS-LR} & {\bf 155} \\
  \: \: $\hookrightarrow$ with accurate selection model ($\chi^2/\nu\leq10$) & 118 \\
 \: \: \: $\hookrightarrow$ with T measurements (incl. fixed $Z = 0.3\,Z_\sun$) & 105 (16) \\
 \:  \: \: \: $\hookrightarrow$ with $L_X$ measurements (incl. fixed $\beta=\frac{2}{3}$) & 104 (8) \\
 \noalign{\smallskip}\hline
\end{tabular}
\tablefoot{Each row in the table represents a subset of the row above it.}
\label{tab:selection}
\end{table}

As our analysis includes sources close to the limits of the sample selection, an accurate model of the selection function is required. As described in Sect.~\ref{sec:xclass}, the selection function model of the parent sample was derived from mock XMM observations. While these simulations were designed to capture the most significant features of a typical X-CLASS field, they did not realistically reproduce unusual pointings or observing conditions. We thus estimated the similarity between a real pointing and one of these idealized simulations, which contain only point sources and a nominal background level.  Our similarity index is the reduced chi-squared value (i.e.,~goodness-of-fit $\chi^2/\nu$) obtained when estimating the background level $\hat{b}$ of a pointing. The value of $\hat{b}$ serves in particular to query the database of selection functions determined from simulated XMM observations with a variety of known background levels $b$ \citep{2012Clerc}. A high value of $\chi^2/\nu$ indicates that an X-CLASS pointing deviates substantially from our set of simulated pointings and that its selection function is not under firm control. This may occur if a large cluster covers a significant portion of the XMM field of view, if anomalies affect the background, or if unusual features impact the observation. Applying a cleaning threshold of $\chi^2/\nu = 10$, set as a compromise between completeness and controlled selection functions, removes 31 X-CLASS fields and 37 X-CLASS-LR objects, leaving 118 groups. One of the advantages of this cleaning procedure is that it does not introduce bias into the analysis of scaling relations. On top of that, we remove 14 clusters that do not have physical parameter measurements. We end up with 104 clusters in our analysis sample. Table \ref{tab:selection} summarizes the sequence of selection steps leading to this sample. We have verified that this supplementary selection results in a similar median mass to the X-CLASS-LR sample.

\subsection{Methodology}

We choose to model the underlying distribution of galaxy clusters with the goal to reproduce the observed luminosity-temperature parameter space. Our approach is similar to the one developed in recent works \citep[e.g.,][]{2016Giles}, and the formalism specifically follows \citet{2022Bahar} prescriptions.

The X-CLASS-LR clusters represent a cluster sub-population of the overall population defined by a mass function and a joint scaling relation between $(L_X, T)$ and mass $M_{500}$.
To fully understand this sub-sample, which was created using several filtering steps, we must model the effects of the various selection criteria applied to the overall X-CLASS selection function. Moreover, we will call $\Lxmes$ and $\Tmes$ the measurements of $L_X$ and $T$, which are noisy values of the intrinsic ones. We assume that the measurement errors on the redshift are negligible because all of our redshifts are spectroscopic redshifts, i.e.,~we identify $\zmes$ with $z$. Finally, we assume that the cosmological parameters are fixed.

We define the joint likelihood of the entire set of observed temperatures and luminosities given our model as the product of individual likelihoods for each cluster (indexed by $i$): 
\begin{equation}
    \mathscr{L}(\Lxmes,\Tmes|I,\theta,\zmes) = \prod_i \mathscr{L}(\Lxmes_i,\Tmes_i|I_i,\theta,\zmes_i)
\end{equation}
Where $\mathscr{L}(\Lxmes_i,\Tmes_i|I_i,\theta,\zmes_i)$ is the likelihood of a single cluster and $\theta$ represents the free parameters of our model :
\begin{equation}
    \mathscr{L}(\Lxmes_i,\Tmes_i|I_i,\theta,\zmes_i) = \frac{P(\Lxmes_i,\Tmes_i,I|\theta,\zmes_i)}{\iint{P(\Lxmes_i,\Tmes_i,I_i|\theta,\zmes_i)}d\Lxmes_i d\Tmes_i}
\end{equation}

The probability that we measure a luminosity $\Lxmes$ and a temperature $\Tmes$ and that the source is included in the sample $I$ with the given parameters $\theta$ and the redshift $\zmes$ can be written as: 
\begin{equation} \label{eq:proba}
    \begin{aligned}
       & P(\Lxmes,\Tmes,I|\theta,\zmes) = \iiint P(\Lxmes,\Tmes|I,L_X,T,M_{500},\theta,\zmes) \\ 
       & \times P(I|L_X,T,M_{500},\theta,\zmes) \times P(L_X,T|M_{500},\theta,\zmes) \times P(M_{500}|\theta,\zmes) \\ 
       & \times dL_X dT dM_{500}
    \end{aligned}
\end{equation}

Where indices are omitted to ease readability.

$P(M_{500}|\theta,\zmes)$ is proportional to the halo mass function at redshift $\zmes$ proposed by \citet{2008Tinker} and computed in our reference cosmology. 

$P(\Lxmes,\Tmes|I,L_X,T,M_{500},\theta,\zmes)$ is a term incorporating the two-dimensional measurement uncertainties. We assume independent and unbiased Gaussian distributions for both $\Lxmes$ and $\Tmes$, with a scale given by the actual measured uncertainty (see Table~\ref{table:measurements}). 

$P(I|L_X,T,M_{500},\theta,\zmes)$ is the probability for a cluster with its physical properties $(L_X,T,M_{500},\zmes)$ to be in our sample with the X-CLASS selection function and our additional selection criteria. The selection function of the X-CLASS survey is given as a function of the true count rate $CR$ and the true core radius of the cluster $R_C$ \citepalias{2021Koulouridis}. We assume a simple relation $R_C=x_c R_{500}$ in order to reflect the expected scaling between gas and total matter spatial distributions in a cluster. In what follows, $x_c$ is left as a free parameter. At this stage, we do not explicitly model the targeting (or pointing) bias inherent to XMM-Newton archival observations, whereby clusters are preferentially located near the pointing center. Finally, we apply an additional size selection to exclude clusters with $R_{500}$ larger than the XMM-Newton field of view.
This leads to an upper value for the mass of a cluster: $M_{500,max}(\zmes)$. The probability $P(I|L_X,T,M_{500},\theta,\zmes)$ depends on $CR \equiv CR(L_X, T, \zmes)$ and $R_C \equiv R_C(M_{500})$ as :

\begin{equation}
    P(I|L_X,T,M_{500},\theta,\zmes) = \mathrm{Sf}\left(CR , R_C , \hat{b} \right) \mathbf{W}\left(0,M_{500,max}(\zmes)\right)
\end{equation}
In this equation, $\mathrm{Sf}$ represents the X-CLASS selection function at the estimated background level $\hat{b}$ of the pointing, and $\mathbf{W}$ is a top-hat function between $0$ and $M_{500,max}$.

$P(L_X,T|M_{500},\theta,\zmes)$ is a joint distribution incorporating the two scaling relations $T-M_{500}$ and $L_X-M_{500}$ as well as the covariance between $L_X$ and $T$ at fixed $M_{500}$. We model the relation $\langle T | M_{500} \rangle$ of the mean temperature $T$ at a given mass $M_{500}$ with a power law dependence on $M_{500}$. We fix the slope and normalization to the values measured in \citet{2015Lovisari} for their total sample. Similarly, the relation of the mean luminosity $L_X$ at a given mass and $M_{500}$ is modeled as follows:
\begin{equation}
    \langle L_X | M_{500} \rangle = C_{L_X} \left(\frac{M_{500}}{M_{\rm piv}}\right)^A .
\end{equation}

The normalization $C_{L_X}$ and the slope $A$ are left as free parameters, and $M_{\rm piv}= 5 \times 10^{13} \ M_{\odot}$ is the pivot in the scaling relation. The narrow redshift range of the X-CLASS-LR sample ($0.07 < \zmes < 0.2$) implies that redshift evolution can be safely neglected. We model the probability distribution $P(L_X,T|M_{500},\theta,\zmes)$ as a two-dimensional multivariate log-normal distribution around the scaling relations $\ln (\langle L_X | M_{500} \rangle(\zmes,M_{500},\theta))$ and $\ln (\langle T | M_{500} \rangle (M_{500}))$. We also take into account intrinsic scatters; specifically, $\sigma_T \equiv \sigma_{\ln T|M_{500}}=0.228$ is fixed \citep[from][originally quoted 
in $\log_{10}$]{2015Lovisari}, while $\sigma_L \equiv \sigma_{\ln L_X|M_{500}}$ is left as a free parameter. Following \citet{2016Mantz}, we account for correlated scatter between $L_X$ and $T$ at fixed mass, with coefficient $\rho=0.56$.

In summary, our model depends on four free parameters: $x_c$ is the relation between the core radius $R_C$ and the radius $R_{500}$, $C_{L_X}$ and $A$ are the normalization and the slope of the $M_{500}-L_X$ relation, and $\sigma_L$ is the intrinsic scatter in $\ln L_X$ at fixed $M_{500}$.

We infer the posterior distribution of model parameters using the MCMC sampler package {\tt emcee} \citep{2013Foreman} with the likelihood described above with 1200 steps of 80 walkers. We take uniform priors in the distributions of $x_c \sim \mathcal{U}(0.005,1)$, $C_{L_X} \sim \mathcal{U}(10^{41},10^{46})$, $A \sim \mathcal{U}(0.5,4)$, and $\sigma_{L} \sim \mathcal{U}(0.01,1.5)$.

\subsection{Inferred model parameter values
\label{sec:results}}

\begin{figure}
    \centering
    \includegraphics[width=0.98\linewidth]{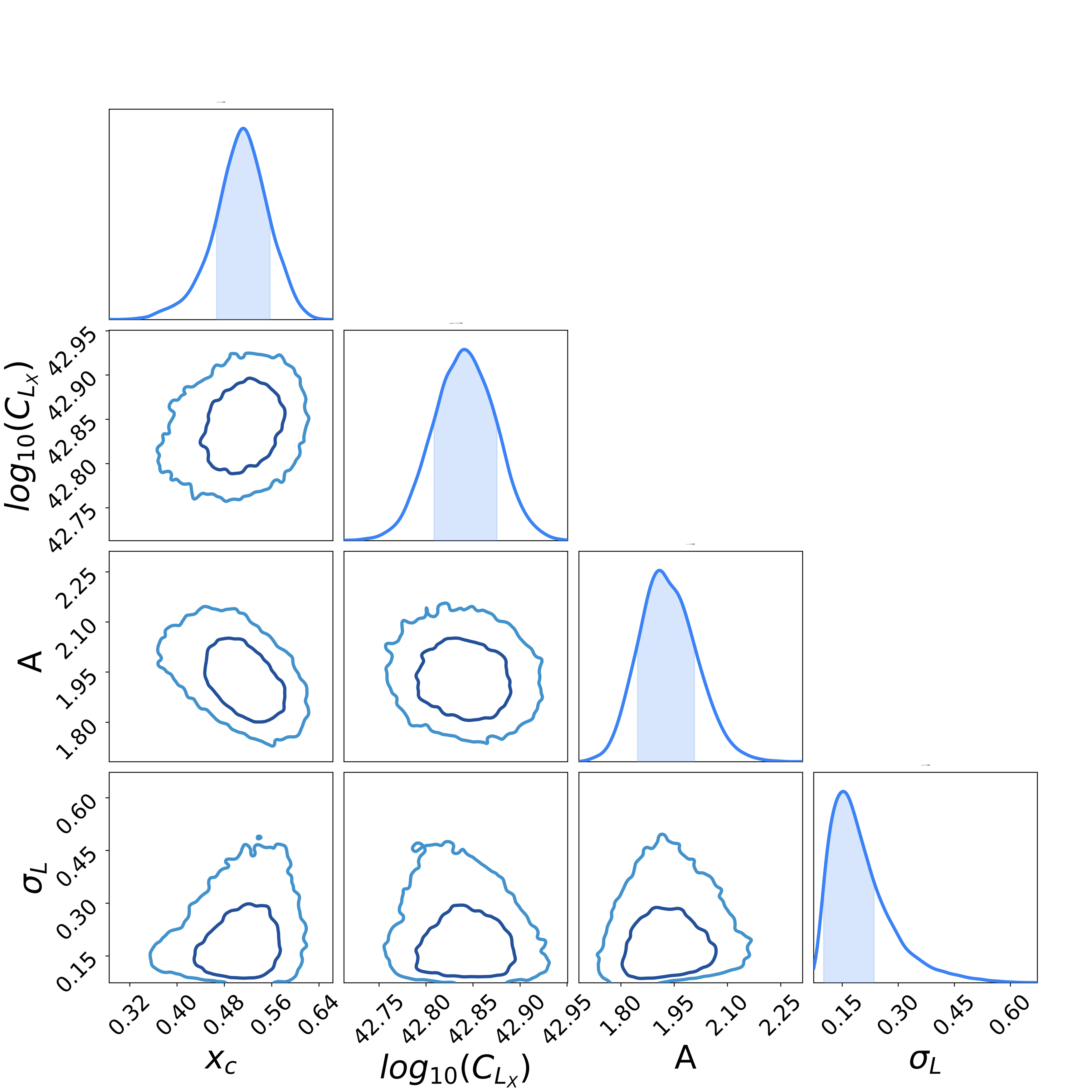}
    \caption{Visualization of the posterior distributions of the free parameters in the model. The inner (outer) contours enclose 68\% (95\%) of the marginalized two-dimensional posterior distributions. Shaded regions display the 68\% interval around the median in the marginalized one-dimensional posteriors.}
    \label{fig:Corner}
\end{figure}

The best fit of the above model gives us the following results for the four free parameters: $x_c=0.51^{+0.05}_{-0.04}$, $\log_{10}(C_{L_X})=42.84^{+0.04}_{-0.03}$, $A=1.9^{+0.10}_{-0.06}$, and $\sigma_L=0.15^{+0.08}_{-0.06}$. The posterior distribution of the parameters allows us to deduce the median and confidence intervals as shown in Fig.~\ref{fig:Corner}. 

The slope $A = 1.9$ is much steeper than the self-similar expectation for the $L_X - M_{500}$ relation ($A=1$). Our result is in agreement within $1\sigma$ with other studies extending to the group regime \citep{2015Anderson, 2015Lovisari, 2025Popesso}. It is also consistent with \citet{2009Pratt}, whose sample includes mostly massive clusters. However, our values of $M_{500}$ derive from the temperature measurements $T$ via a rather general relation for X-ray clusters, because we do not have direct access to the total mass. Consequently, our free parameters $A$, $C_{L_X}$, and $\sigma_L$ should be considered as intermediate parameters that we use to derive the $L_X-T$ relation in Section \ref{sec:Lx-T}. These intermediate parameters are needed to incorporate the selection function in the model.

The value of the intrinsic scatter in luminosity at fixed mass ($\sigma_L$) is rather small compared to that reported in other studies \citep[see, for example,][]{2011Eckmiller, 2016Mantz}, where significantly higher values lie approximately $3\sigma$ above our result. One possible explanation is the difference in the redshift ranges considered: while \citet{2015Lovisari} focus on low-redshift clusters ($0.01\leq z \leq 0.035$), it has been shown by \citet{2023Damsted} that the intrinsic scatter in X-ray luminosity increases sharply at very low redshifts ($z\leq0.15$). Therefore, our results are more consistent with measurements at higher redshifts. Another possible explanation is that the selection function is contaminated by the possible presence of a central AGN in some clusters. \citet{2014Clerc} showed that the brighter the central AGN of a cluster is relative to the ICM flux, the less likely the source is to be classified as extended. Therefore, our sample may systematically lack a sub-population of low surface brightness AGN-contaminated clusters, which could result in the inferred intrinsic scatter being biased towards lower values. The low value of $\sigma_L$ could also be explained by a too high value of $\sigma_T$ in our model. To test this hypothesis, we ran our analysis with a smaller value of the scatter in $\ln T$, $\sigma_T=0.1$. We end up with a significantly lower normalization $\log_{10}(C_{L_X})=42.63^{+0.05}_{-0.03}$ and a significantly higher intrinsic scatter in $\ln L_X$, $\sigma_L=0.76^{+0.05}_{-0.06}$. This latter value is closer to the values reported in other studies. However, we report no significant change in the slope $A$ and $x_c$.

Our model assumes a uniform value of $x_c$ in the relation $R_C=x_c R_{500}$ for all our clusters. An evolution of $x_c$ with redshift or total mass could introduce a systematic effect that contributes to the observed discrepancy. In the literature, only a few studies allow the parameter $x_c$ to vary freely \citep{2012Clerc}, while most analyses fix it to values around 0.1-0.2 \citep{2016Giles}. We find a significantly larger value of $x_c$ than \citet{2012Clerc}, who report $x_c=0.24\pm0.04$ using the entire X-CLASS sample. This could suggest that at fixed mass, the central gas distribution in low-mass galaxy clusters is more extended than typically assumed. However, our constraint is an indirect one, since the radial extent of clusters only enters through the selection function in our model.

We also tested the influence of the correlation between $L_X$ and $T$ at fixed $M_{500}$ by running our model with two different values obtained from different analyses: $\rho=0.56$ \citep{2016Mantz} and $\rho=-0.06$ \citep{2016bMantz}. We find that with a lower (and negative) correlation, we end up with a lower normalization $C_{L_X}$ and a higher slope $A$.
This behavior can be understood in terms of selection near the detection threshold. If we take a cluster with a fixed mass $M_{500}$ close to the detection limit, a higher correlation means that if it is detected, it will appear brighter and hotter. A correlation close to 0, on the other hand, means that it will only appear brighter and therefore lower the $\langle L_X | M_{500} \rangle$ for the low-mass clusters. This degeneracy between $\rho$, $A$, and $C_{L_X}$ is intrinsic to our dataset and cannot be raised without additional observations.

\subsection{The $L_X-T$ scaling relation}
\label{sec:Lx-T}

\begin{figure*}
    \centering
    \begin{minipage}{0.7\textwidth}
        \includegraphics[width=\linewidth]{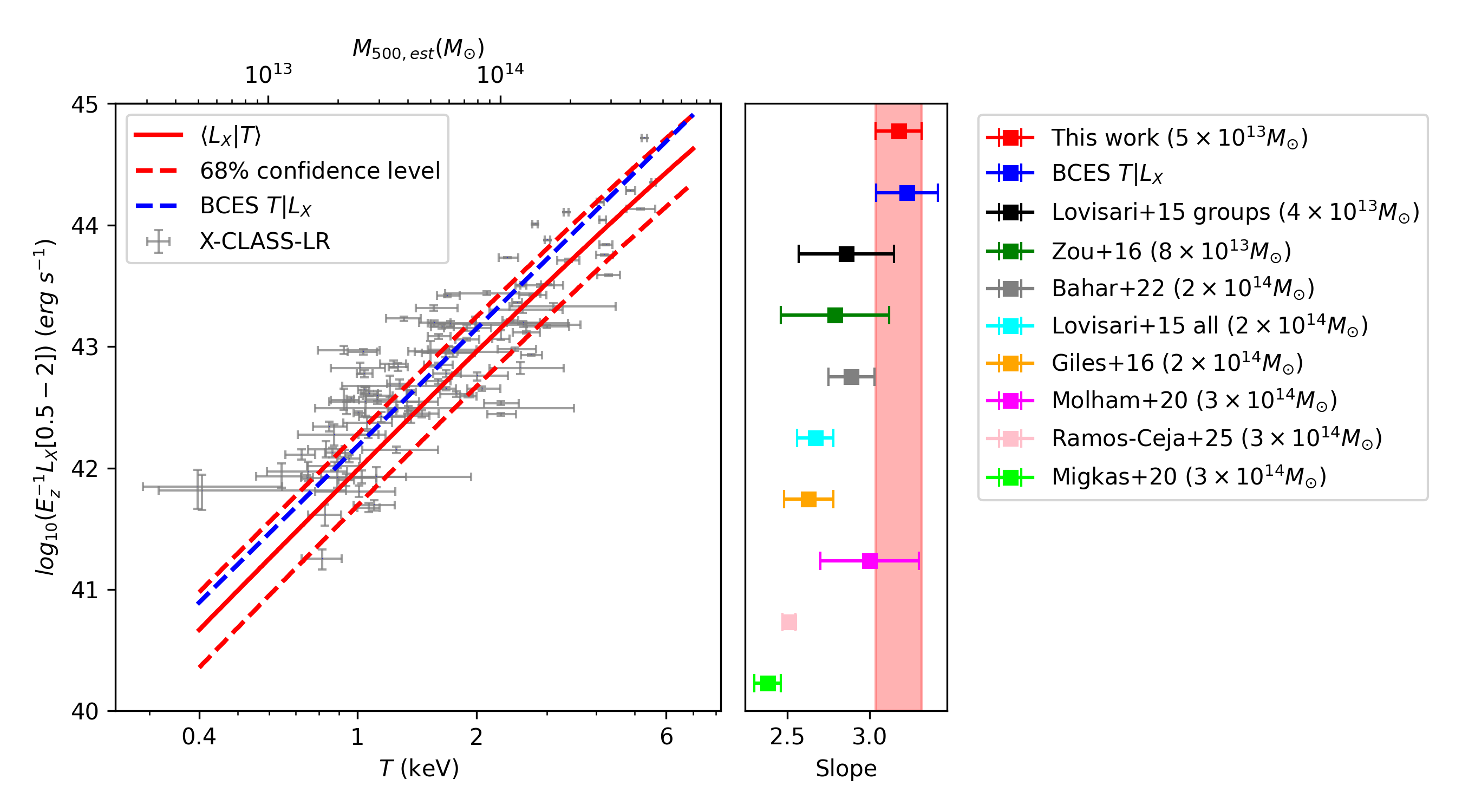}
    \end{minipage}
    \hfill
    \begin{minipage}{0.25\textwidth}
        \caption{Left: $L_X-T$ scaling relation of the 104 X-CLASS-LR clusters. The blue dashed line represents a fit using the BCES regression method \citep{1996Akritas, 2012Nemmen}, while the red line is the best fit result from the full simulation of the likelihood function. The $1 \sigma$ errors are plotted as dashed red lines. Right: Comparison of the slope for the $L_X-T$ scaling relation with other X-ray selected samples. The median of the estimated $M_{500}$ of each sample is indicated in parentheses.}
    \end{minipage}
    \label{fig:Lx-T}
\end{figure*}

The final step in the above process consists in establishing the $L_X-T$ relation. Our preliminary approach uses a simple BCES X|Y regression algorithm \citep{1996Akritas, 2012Nemmen}, although doing so does not take into account the complex selection function of X-CLASS. Nevertheless, this can guide the rest of the study, and we obtained a slope $B_{\rm BCES}=3.3\pm 0.2$ and a normalization $\log_{10} (D_{\rm BCES})=43.15\pm 0.04$ using the standard parametrization:
\begin{equation}
    \langle L_X|T \rangle = D \left(  \frac{T}{T_{\rm piv}} \right)^B
\end{equation}
Where $T_{\rm piv}=2$\,keV is the pivot point.

The more detailed and complete approach set up in our model enables us to recover the above relation using a set of parameters $\theta_c$. The following equations apply:
\begin{equation}\label{eq:p_lx_t}
    P(L_X|T,\theta_c,\zmes) = \frac{ P(L_X,T|\theta_c,\zmes) } {\int_0^{\infty} P(L_X,T|\theta_c,\zmes) dL_X }
\end{equation}

\begin{equation}
    P(L_X,T|\theta_c,\zmes) = \int_0^{\infty} P(L_X,T|M_{500},\theta_c,\zmes) P(M_{500}|\theta_c,\zmes) dM_{500}
\end{equation}

To propagate the uncertainties on the free parameters $\theta$, we computed this probability (left-hand side of Eq.~\ref{eq:p_lx_t}) for each of the last 2000 steps of the MCMC chains on a grid of temperatures. We then sampled the resulting distributions. We combined the luminosity samples for each temperature along the grid to obtain a final sampling of $P(L_X|T,\theta,\zmes)$. 
 This procedure also delivers the intrinsic dispersion in $\ln(L_X)$ at each fixed temperature, which takes value $\sim 0.66$ (i.e.,~0.29\,dex, dashed lines in Fig.~\ref{fig:Lx-T}).
To derive the normalization and the slope of the $L_X-T$ relation (and their uncertainties), we performed a linear fit of the ($L_X$, $T$) samples computed at each of the last 2000 steps. This yields a distribution of values for normalization and slope. Each distribution is modeled with a Gaussian to obtain an average slope $B=3.2\pm 0.1$ and a normalization $\log_{10}(D)=42.97\pm0.03$.

Again we find a steeper slope than the self-similar expectation ($L_X \propto T^{3/2}$), in agreement with recent studies. Our derived intrinsic scatter also agrees with those studies.

The right panel of Fig. \ref{fig:Lx-T} compares our best-fit slope of the $L_X-T$ scaling relation with results from other X-ray selected samples. Our slope is consistent within $\sim1\sigma$ with \citet{2015Lovisari} (group sub-sample, [0.1-2.4] keV band), \citet{2016Zou}, and \citet{2022Bahar}, which all focused on low-mass systems, although we find a slightly higher slope. We are also consistent with the result from \citet{2020Molham}, who worked on higher-mass X-CLASS clusters (only 8 clusters are present in both their sample and ours). However, they argue that their result may be biased at the few-percent level towards high values due to the effect of the Malmquist bias.
In contrast, our slope is significantly steeper than those reported in studies dominated by higher-mass clusters \citep{2016Giles, 2020Migkas, 2025Ramos-Ceja}.
We can also observe a trend towards steeper slope as the median of the estimated mass of the sample decreases.
Several works have shown that the $L_X-T$ scaling relation is steeper for galaxy groups than for galaxy clusters \citep{2012Maughan, 2021Lovisari}. This behavior may be attributed to the shallower potential wells of galaxy groups, making them more sensitive to non-gravitational heating processes such as AGN feedback, supernovae feedback, and mergers \citep{2022Lovisari}.
A similar trend is observed in cosmological hydrodynamic simulations. For example, the Millennium simulation \citep{2016Chon} shows that including AGN feedback reduces the luminosity of low-mass systems, bringing the predicted scaling relation into better agreement with observations. Comparable behavior was also observed in the SIMBA simulation \citep{2020Robson}, which predicts a break in the power law of the $L_X-T$ relation towards a steeper slope at $T \lesssim 1~keV$. These simulation results therefore support the interpretation that the steeper slope observed in low-mass systems is driven by the stronger impact of feedback processes in shallow potential wells.
Therefore, the discrepancy with high-mass studies may simply reflect the higher fraction of galaxy groups in our sample. This higher group fraction could also account for our steeper slope compared with the other low-mass studies: $\sim 69\%$ of our clusters are groups with masses below $10^{14}M_{\odot}$, a figure to be compared with \citet{2022Bahar}, for example, reporting only $26\%$.
Finally, the significantly larger value of $x_c$ found in our analysis may also reflect the impact of central AGN feedback, which can redistribute gas to larger radii, thereby increasing $x_c$ and potentially contributing to the steepening of the $L_X-T$ relation.

The choice of a fixed $T-M_{500}$ scaling relation could have an impact on the final determination of the slope of the $L_X-T$ relation. We selected the $T-M_{500}$ from \citet{2015Lovisari} because they are among the few studying this relation for low-mass clusters. We tested the influence of the intrinsic scatter of $T$ at fixed $M_{500}$ by using our run with $\sigma_T=0.1$ and then recovering the $L_X-T$ relation. Finally, we end up with a normalization $\log_{10}(D)=42.97\pm0.05$ and a slope $B=3.3\pm0.1$, consistent with our main model. So the value of $\sigma_T$ only has an impact on the parameters $C_{L_X}$ and $\sigma_L$ of the model.
This discussion highlights the intermediate nature of our free parameters ($A$, $C_{L_X}$, and $\sigma_L$), whose determination is intimately linked to our assumption of the $T-M_{500}$ relation.
Moreover, we repeated all our measurements and defined our model using the $T-M_{500}$ relation from the bright cluster sample of the XXL survey \citep{2016Lieu}, and we ended up with a normalization $\log_{10}(D)=42.96\pm0.04$ and a slope $B=3.2\pm0.1$. The results are consistent with our main model, and this reinforces our confidence in the hypothesis of the measurements and the model developed in this paper.

\section{Summary \label{sec:conclusion}}
In order to propose and construct a sample of low redshift groups and clusters, we first worked on the X-CLASS survey database to update the redshift estimates, which are essential for analyzing the properties of the sample. Photometric redshifts were built using two independent approaches combined to submit a robust photo-z. They were validated and implemented for 87\%\ of the clusters with no spectroscopic redshift in the initial database \citepalias{2021Koulouridis}, taking advantage of the fully released DR10 data. The data are made publicly available in the X-CLASS database, as well as updated redshift information for many clusters. We provide a new classification scheme for the redshift information also available in the database, which is completed as much as possible for scientific purposes.
Moreover, for low redshift clusters located in the Northern hemisphere, we completed the sample with spectroscopic determination of all X-CLASS clusters at redshift smaller than 0.2 thanks to a dedicated survey with the MISTRAL instrument of the T193 telescope at the Observatoire de Haute-Provence. By construction of X-CLASS, most of the sources of this low redshift sample (X-CLASS-LR) are low-mass clusters and groups of galaxies. With 155 members, the X-CLASS-LR sample represents, in this mass and redshift range, the largest sample with high-quality measurements. 

Such additional X-ray measurements have been obtained for the clusters in the sample thanks to a detailed spatial and spectral analysis of the X-ray data. The X-ray hot gas luminosity and temperature are determined for nearly all clusters, as well as the $R_{500}$ radius. When the quality of the data permits, the metallicity is also determined by the fit; otherwise, it is fixed to the canonical value. All these physical measurements are used to model the $L_X - T$ scaling relation, taking into account the most relevant selection effects involved in creating the sample and the different steps in the measurement procedure. The $L_X - T$ relation is finally fit with a standard power law and compared to previous results:
\begin{equation}
    \langle L_X|T \rangle = 10^{42.97\pm0.03} \left(  \frac{T}{2 {\rm keV}} \right)^{3.2 \pm 0.1} \;\;\;;\;\; \sigma_{\ln L_X | T} \approx 0.66 
\end{equation}

The slope $B=3.2 \pm 0.1$ we determine is steeper than that published in previous analyses, particularly those carried out with eROSITA data. Our results are consistent with those produced on similar samples as ours, i.e., low-mass clusters and groups, whereas for more massive samples, the slope of the relationship takes a smaller value. This result can be interpreted as the outcome of stronger baryonic processes in low-mass groups and clusters, whose weaker gravitational potential does not retain the expelled gas, in particular as a consequence of feedback from a putative AGN in the center. We thus expect a larger extent of the X-ray emission and a lower luminosity at a given temperature. We propose this preliminary interpretation, which requires further study with larger samples and deeper observations. The increasing size of samples selected in X-rays (XMM-Newton archive, eROSITA) and at other wavelengths (e.g.,~by Sunyaev-Zel'dovich effect and in the optical bands) offers promising perspectives on this matter.

\section*{Data availability}

Tables~\ref{table:redshifts}, \ref{table:cluster_redshifts}, \ref{table:fullredshifts}, and \ref{table:measurements} are only available in electronic form at the CDS via anonymous ftp to cdsarc.u-strasbg.fr (130.79.128.5) or via http://cdsweb.u-strasbg.fr/cgi-bin/qcat?J/A+A/.

\begin{acknowledgements}
We would like to thank the person who carefully reviewed our article and suggested significant improvements that helped strengthen the impact of our findings. 
We thank M.~Kluge for useful discussions on the derivation of cluster photometric redshifts and J.~Sanders for the development of the MBPROJ2D software.

This work was supported by the "Programme National Cosmology et Galaxies" (PNCG) of CNRS/INSU with INP and IN2P3, co-funded by CEA and CNES.
\\
The Legacy Surveys consist of three individual and complementary projects: the Dark Energy Camera Legacy Survey (DECaLS; Proposal ID \#2014B-0404; PIs: David Schlegel and Arjun Dey), the Beijing-Arizona Sky Survey (BASS; NOAO Prop. ID \#2015A-0801; PIs: Zhou Xu and Xiaohui Fan), and the Mayall z-band Legacy Survey (MzLS; Prop. ID \#2016A-0453; PI: Arjun Dey). DECaLS, BASS and MzLS together include data obtained, respectively, at the Blanco telescope, Cerro Tololo Inter-American Observatory, NSF’s NOIRLab; the Bok telescope, Steward Observatory, University of Arizona; and the Mayall telescope, Kitt Peak National Observatory, NOIRLab. Pipeline processing and analyses of the data were supported by NOIRLab and the Lawrence Berkeley National Laboratory (LBNL). The Legacy Surveys project is honored to be permitted to conduct astronomical research on Iolkam Du’ag (Kitt Peak), a mountain with particular significance to the Tohono O’odham Nation.

NOIRLab is operated by the Association of Universities for Research in Astronomy (AURA) under a cooperative agreement with the National Science Foundation. LBNL is managed by the Regents of the University of California under contract to the U.S. Department of Energy.

This project used data obtained with the Dark Energy Camera (DECam), which was constructed by the Dark Energy Survey (DES) collaboration. Funding for the DES Projects has been provided by the U.S. Department of Energy, the U.S. National Science Foundation, the Ministry of Science and Education of Spain, the Science and Technology Facilities Council of the United Kingdom, the Higher Education Funding Council for England, the National Center for Supercomputing Applications at the University of Illinois at Urbana-Champaign, the Kavli Institute of Cosmological Physics at the University of Chicago, Center for Cosmology and Astro-Particle Physics at the Ohio State University, the Mitchell Institute for Fundamental Physics and Astronomy at Texas A\&M University, Financiadora de Estudos e Projetos, Fundacao Carlos Chagas Filho de Amparo, Financiadora de Estudos e Projetos, Fundacao Carlos Chagas Filho de Amparo a Pesquisa do Estado do Rio de Janeiro, Conselho Nacional de Desenvolvimento Cientifico e Tecnologico and the Ministerio da Ciencia, Tecnologia e Inovacao, the Deutsche Forschungsgemeinschaft and the Collaborating Institutions in the Dark Energy Survey. The Collaborating Institutions are Argonne National Laboratory, the University of California at Santa Cruz, the University of Cambridge, Centro de Investigaciones Energeticas, Medioambientales y Tecnologicas-Madrid, the University of Chicago, University College London, the DES-Brazil Consortium, the University of Edinburgh, the Eidgenossische Technische Hochschule (ETH) Zurich, Fermi National Accelerator Laboratory, the University of Illinois at Urbana-Champaign, the Institut de Ciencies de l’Espai (IEEC/CSIC), the Institut de Fisica d’Altes Energies, Lawrence Berkeley National Laboratory, the Ludwig Maximilians Universitat Munchen and the associated Excellence Cluster Universe, the University of Michigan, NSF’s NOIRLab, the University of Nottingham, the Ohio State University, the University of Pennsylvania, the University of Portsmouth, SLAC National Accelerator Laboratory, Stanford University, the University of Sussex, and Texas A\&M University.

BASS is a key project of the Telescope Access Program (TAP), which has been funded by the National Astronomical Observatories of China, the Chinese Academy of Sciences (the Strategic Priority Research Program “The Emergence of Cosmological Structures” Grant \# XDB09000000), and the Special Fund for Astronomy from the Ministry of Finance. The BASS is also supported by the External Cooperation Program of Chinese Academy of Sciences (Grant \# 114A11KYSB20160057), and Chinese National Natural Science Foundation (Grant \# 12120101003, \# 11433005).

The Legacy Survey team makes use of data products from the Near-Earth Object Wide-field Infrared Survey Explorer (NEOWISE), which is a project of the Jet Propulsion Laboratory/California Institute of Technology. NEOWISE is funded by the National Aeronautics and Space Administration.

The Legacy Surveys imaging of the DESI footprint is supported by the Director, Office of Science, Office of High Energy Physics of the U.S. Department of Energy under Contract No. DE-AC02-05CH1123, by the National Energy Research Scientific Computing Center, a DOE Office of Science User Facility under the same contract; and by the U.S. National Science Foundation, Division of Astronomical Sciences under Contract No. AST-0950945 to NOAO.

The Photometric Redshifts for the Legacy Surveys (PRLS) catalog used in this paper was produced thanks to funding from the U.S. Department of Energy Office of Science, Office of High Energy Physics via grant DE-SC0007914.
\end{acknowledgements}

\bibliographystyle{aa}
\bibliography{biblio}

\begin{appendix}

\section{Measuring photometric redshifts of X-CLASS clusters} 
\label{sec:photoz-annex}
The following presents the two methods used to measure the photometric redshifts of X-CLASS clusters. In particular, we detail the {\tt photXclus} software that we developed specifically for this purpose, for which this is the first formal description. We provide a more concise presentation of {\tt RedMaPPer} and refer the reader to \citet{2020IderChitham} and \citet{2024Kluge} for a detailed description.

\subsection{{\tt photXclus}}
\label{sec:photXclus}
We developed the {\tt photXclus} algorithm to estimate the photometric redshift and richness of X-CLASS clusters using catalogues of galaxy photometric redshifts in the DESI Legacy Survey DR9 \citep{2021Zhou}. This method treats the problem of estimating the cluster redshift as a Poisson process over two populations of galaxies (cluster members and foreground/background galaxies), enabling the cluster redshift and richness to be estimated jointly.

For any galaxy observed in the field of view of an X-ray-detected cluster, we assume it is either (i) a cluster galaxy, or (ii) a foreground or background field galaxy. To obtain a joint estimate of the redshift and richness, we also consider a control region in which we assume that only field galaxies are observed.

Having assumed a model for the redshift and radial distributions of each population, the posterior of model parameters $\theta$ given data $d$ can be written \citep[see][for a similar approach in a different context]{2005Andreon}:

\begin{equation}
p(\theta | d) = p(\theta) \prod_{\underset{j}{\rm datasets}} \left[ \prod_{\underset{i}{\rm galaxies}} p(d_{ij} | \theta) \right] e^{-N_{{\rm exp},j}},
\end{equation}
\noindent where ${\rm datasets} = \{{\rm cluster}, {\rm control}\}$. For a given galaxy $i$ in a dataset $j$, the \textit{extended} likelihood $p (d_{ij} | \theta)$ can be written as :
\begin{equation}
p(d_{ij} | \theta) = \delta_j \times \lambda_{\rm c} \times p_{\rm clus} + n_{{\rm gal}, f} \times \Omega_j \times p_{\rm field},
\end{equation}\label{eq:likelihood}
\noindent where $\delta_j$ is one for cluster regions and 0 for control regions, $\lambda_{\rm c}$ is the number of cluster galaxies (richness), $n_{{\rm gal}, f}$ is the field galaxy density, and $\Omega_j$ is the area of either the cluster line-of-sight (LOS) or the control field-of-view. We write $d_{ij} = (z_{ij}, r_{ij})$, with $z_{ij}$ the galaxy redshift and $r_{ij}$ the projected distance from the cluster center.

\begin{equation}
N_{\rm exp} = \int_{z_{\rm inf}}^{z_{\rm sup}} \int_{0}^{R_{\rm fit}} p(z, r | \theta)~dr dz
\end{equation}
is the expected number of galaxies given the model parameters $\theta$. This term in the posterior distribution comes from modeling galaxies as a Poisson point process, with the number of galaxies in the dataset being a model parameter.

To ensure high completeness up to $z = 1.5$ and relatively stable expected number counts across the field over the considered redshift range, we select bright galaxies using the characteristic luminosity $m^*(z)$ as a reference. It corresponds to the redshift-dependant, passively evolving knee of the Schechter luminosity function, computed using {\tt easyGalaxy} \citep{Mancone2012}. We calibrated it with the characteristic absolute magnitude $M_i^*$ measured in the SDSS $i$ band at $z = 0.1$ and extracted from the galaxy luminosity functions fit of \citet{Ramos2011} in the CFHTLS. In practice, we retain only galaxies with magnitudes in the $r$ and $z$ bands satisfying $m < m^*(z) + 1$.

The cluster is modeled with a fixed radial profile and center, while the cluster redshift $z_c$ and richness $\lambda_{c}$ are the model parameters.
\begin{equation}
\begin{split}
   p_{\rm clus}(z, r) &= p_{\rm clus}(z) \times p_{\rm clus}(r)\\
                    &= \mathcal{N}(z | z_{\rm clus},~\sigma_z^{\rm gal}(z_{\rm clus})) \times {\rm Plummer}(r, R_{\rm fit}),
\end{split}
\end{equation}
where $z_{\rm clus}$ is the cluster redshift, $\sigma_z^{\rm gal}(z_{\rm clus})$ is the expected individual galaxy photometric redshift uncertainty at the cluster redshift and $R_{\rm fit}$ is the cluster radius fitted by the XCLASS pipeline. $\mathcal{N}$ represents a normal distribution. The radial galaxy distribution is taken as a Plummer profile \citep{2012Ascaso}.
The field is modeled as a spatially uniform distribution and described as a piecewise constant function in redshift space.
\begin{equation}
\begin{split}
   p_{\rm field}(z, r) &= p_{\rm field}(z) \times p_{\rm field}(r)\\
                    &= \sum_{k=1}^{N} p_k \, \mathbb{I}(z^{\rm bins}_{k} \leq z < z^{\rm bins}_{k+1}) \times p_{\rm field}(r), 
\end{split}
\end{equation}
where $ \mathbb{I}(z^{\rm bins}_{k} \leq z < z^{\rm bins}_{k+1})$ is an indicator function that takes the value 1 when $z$ is within the $k$-th bin, and 0 otherwise, and
\begin{equation}
p_{\rm field}(r) =
\begin{cases}
2\pi r / \Omega_{j} & \text{in cluster regions} \\
1 & \text{in control regions}.
\end{cases}
\end{equation}

Finally, we estimate the posterior distribution of the parameter set $\vect \theta = \left ( z_{\rm clus}, \lambda_{\rm clus}, n_{\rm gal, f}, \{p_k\}_{k = 1 \cdots N_{\rm bins}}\right )$ knowing the data $\vect d = \left ( \{z_{ij}, r_{ij} \}_{i = 1 \cdots N_{\rm galaxies}; j = 1 \cdots N_{\rm dataset}}\right )$. 

For all parameters but $z_{\rm clus}$, we use flat, uninformative priors. For $p(z_{\rm clus})$ we use information from the X-ray detection of the cluster coupled with galaxy cluster expected properties to build an informative prior.

In particular, we use the cluster's typical size, $R_{\rm fit},$ measured by the X-CLASS pipeline, to deduce a prior on its redshift. Clusters that appear as strong X-ray emissions in the XMM archive are more likely to be at a lower redshift. For a given $R_{\rm fit}$ measured in arcminutes, we convert the radius to a physical size at a given redshift, $z$. If the implied physical size is larger than $2$ Mpc, the probability of observing a cluster at this redshift declines according to a half-normal distribution centered at the redshift where $R_{\rm fit} = 2$ Mpc. 
A physical radius of $2$ Mpc corresponds to the upper end of the observed cluster size distribution (comparable to $R_{200}$ for the most massive systems), and clusters larger than this are rare according to both observations and halo mass function predictions. In principle, a symmetrical prior could be applied to set a minimum size. However, since we do not know the mass function of the X-CLASS sample, distinguishing between small, low-redshift clusters and massive, high-redshift clusters is a degenerate problem. Therefore, we decided to be conservative and not set a prior on the minimum allowed size at a given redshift. The prior can be written as follows:
\begin{equation}
p(z_{\rm clus}) = \begin{cases} 
      1 & r_{\rm phys}(z_{\rm clus}) \leq 2 {\rm Mpc} \\
      \exp\left(- \frac{(z_{\rm clus} - z_{2 {\rm Mpc}})^2}{2 \sigma^2}\right) &  r_{\rm phys}(z_{\rm clus}) > 2 {\rm Mpc}
      \end{cases} 
\end{equation}

We draw 51000 samples from the posterior distribution using MCMC with the {\tt zeus} package \citep{2021Karamanis} using 68 chains. Initial parameters in each chain are chosen following a uniform distribution in $0 < z < 1.5$ for $z_{\rm clus}$ and a Poisson distribution around the observed values for $\lambda_{\rm clus}$, $n_{\rm gal, f}$, and $\{n_{\rm gal, f} \times p_k\}_{k = 1 \cdots N_{\rm bins}}$.

We detect peaks in the two-dimensional posterior marginal $p(z_{\rm clus}, \lambda_{\rm clus})$ using persistence homology \citep[see e.g.,][]{2011Sousbie}, assign each cluster its $z_{\rm clus}$, and $\lambda_{\rm clus}$ at the peak with highest persistence, and compute the 68\% credible interval of the posterior around the peak. In this work, we only keep estimated redshifts for which the 68\% credible interval on redshift is smaller than 0.15.

\subsection{{\tt RedMaPPer}}\label{sec:eROMaPPer}
In this study, {\tt RedMaPPer} was run in scanning mode using the X-ray central position as the prior position. In this setup, the richness and redshift of each cluster are estimated at the X-ray central position using galaxies within a scaling radius $R_\lambda = 1.0 h_{70}^{-1} {\rm Mpc} (\lambda/100)^{0.2}$. 

In practice, the richness parameter $\lambda$ is first evaluated using a fixed initial redshift grid. Then, the cluster redshift is taken as the one that maximizes the likelihood:
\begin{equation}
\ln \mathcal{L}_\lambda = - \frac{\lambda}{S} - \sum_i\ln(1 - p_{{\rm mem}, i}).
\end{equation}

Starting from the initial richness and redshift estimates, the redshift estimate is iteratively refined using the algorithm described in Section 7.2 of \citet{2014Rykoff}, in which the likelihood of observing the galaxy colors at a given redshift estimate is maximized until convergence.

The {\tt RedMaPPer} version had its red-sequence model calibrated using a large sample of publicly available, spectroscopically confirmed cluster catalogs, as described in detail in \citet{2024Kluge}.

\section{MISTRAL individual redshifts} \label{sec:zmistral}
This section lists all the galaxies observed during the MISTRAL runs for which a redshift was determined. All stars detected in the slits have been removed. We used data from the LS DR9 or DR10, as well as data from ESASky and PAN-STARRS1 images, which are available online, to identify and locate the galaxies. The data are presented in Table~\ref{table:redshifts}.

\begin{table*}
\centering
\caption{\label{table:redshifts} Results of the MISTRAL observing runs at OHP.}
	\begin{tabular}{lrrcccccl}
		\hline\hline\noalign{\smallskip}
        X-CLASS & R.A.  & DEC.  & $g$ & $r$ & zphot & $z$ & Conf &  Comments \\
	    & (deg) & (deg) & & & & & & \\
        \noalign{\smallskip}\hline\noalign{\smallskip}

xclass0297\_G1  & 68.5873	  & -8.4908 & 18.27 & 16.73  & 0.247 & 0.2392 & 3 & BCG  \\
xclass0297\_G2  & 68.5703	  & -8.5124 & 19.10 & 17.52  & 0.250 & 0.2397 & 3 &      \\
xclass0297\_G3  & 68.5564	  & -8.5095 & 20.02 & 18.54  & 0.232 & 0.2404 & 3 &      \\
xclass0297\_G4  & 68.5681	  & -8.5067 & 19.86 & 18.53  & 0.224 & 0.2406 & 2 &      \\
xclass0473\_G1  & 7.4357   &  4.8736 & 18.06	& 16.78  & 0.205 & 0.2069  & 3 &     \\
xclass0473\_G2  & 7.4408   &  4.8529 & 18.80  & 17.53  & 0.209 & 0.2050  & 2  &      \\
...  & ... & ... & ... & ... & ... & ... & ... & ... \\
\noalign{\smallskip}\hline
	\end{tabular}
	\tablefoot{For each X-CLASS cluster (1), the galaxies are listed with their coordinates (2,3). The magnitudes in $g$ and $r$ are derived from the DESI Legacy Survey when available (4,5), as is the photometric redshift (6). The confidence level (7) ranges from 1 (tentative, no more than 2 absorption lines clearly identified) to 3 (good, based on a minimum of 3 strong absorption lines). Some comments are added, particularly the identification of the BCG when possible (8). Only part of the table is presented here. The full table is available at the CDS.}
\end{table*}

\section{Updated X-CLASS redshifts\label{sec:redshiftable}}

This section provides details on the final redshift evaluation for all X-CLASS clusters. We first present the final table of all new redshifts validated in the MISTRAL sample (Table~\ref{table:cluster_redshifts}). Clusters with a redshift smaller than 0.2 are included in the X-CLASS-LR sample, and the others are simply updated in the new database presented below.

\begin{table}[ht]
\caption{Redshifts of the 52 X-CLASS clusters measured with MISTRAL.}
\label{table:cluster_redshifts}
\centering 
\begin{tabular}{l r r c c}
\hline\hline\noalign{\smallskip}
X-CLASS & RA & DEC & $z$ & S \\
Id & (deg) & (deg) & & (flag) \\
\noalign{\smallskip}\hline\noalign{\smallskip}
0297             & 68.5703	 & --8.5124 & 0.240       & 5    \\
0473             & 7.4357	 &   4.8736 & 0.206       & 5    \\
0677             & 56.5638   & 24.30241 & 0.215       & 5    \\
0721             & 95.5470   & 78.3089  & 0.144       & 5    \\
0788             & 196.3651  & 67.58920 & 0.423       & 2    \\
1032             & 149.8855  & 5.4326   & 0.214       & 5    \\
1193             & 195.4398  & 58.9528  & 0.239       & 5    \\
1240             & 233.5705  & 73.4873  & 0.119       & 5    \\
1330             & 212.0500  & 78.6110  & 0.196       & 5    \\
1345             & 125.3754  & 1.0664   & 0.086       & 5    \\
1365             & 155.5969  & 22.0398  & 0.157       & 5    \\
1494             & 14.6666   & 38.1789  & 0.222       & 4    \\
1495             & 14.3348   & 37.9770  & 0.242       & 5    \\
1653             & 357.9518	 & 20.1090  & 0.145       & 4    \\
1655             & 202.6086  & 24.23220 & 0.462       & 4    \\
1660             & 11.9622   & 85.45596 & 0.280       & 4    \\
1745             & 221.0436  & 77.1682  & 0.473       & 2    \\
1774             & 29.3125   & 37.9487  & 0.316       & 3    \\
1805             & 139.5906  & 69.0165  & 0.117       & 3    \\
2045             & 175.0631  & 2.9424   & 0.215       & 5    \\
2249             & 125.1272	 & 20.934   & 0.264       & 2    \\
2345             & 162.0686  & --4.2093 & 0.246       & 4    \\
2355             & 180.4070  & 30.0037  & 0.164       & 2    \\
2373             & 22.0862   & 29.2415  & 0.053       & 2    \\
2482             & 253.0800  & 55.5583  & 0.323       & 4    \\
2483             & 264.5287  & 60.1050  & 0.331       & 2    \\
2611             & 16.9167   & 15.2555  & 0.187       & 5    \\
2638             & 356.8598	 & 15.5585  & 0.200       & 4   \\
2702             & 250.6249  & 65.6167  & 0.182       & 5    \\
2730             & 274.2379  & 42.7754  & 0.241       & 4    \\
2838             & 52.3021   & 5.3089   & 0.083       & 5    \\
2889             & 314.7862	 & 5.1588   & 0.108       & 4    \\
2908             & 4.9502	 & 3.6952   & 0.268       & 4    \\
3038             & 124.6218  & 70.7924  & 0.205       & 5    \\
3054             & 19.7576   & 3.394    & 0.400       & 5    \\
3135             & 30.3431   & --2.3288 & 0.238       & 5    \\
3299             & 109.2039  & 37.6766  & 0.069       & 5    \\
3353             & 255.5798  & --0.9795 & 0.091       & 4    \\
3447             & 53.8294	 & --6.9401 & 0.220       & 5    \\
3452             & 37.2242   & 31.7493  & 0.058       & 5    \\
3453             & 37.2480   & 31.5549  & 0.342       & 2    \\
3454             & 36.7519   & 20.5067  & 0.268       & 4    \\
21432            & 117.4894  & 55.8644  & 0.173       & 5    \\
21544            & 215.7493  & 19.5934  & 0.270       & 4    \\
21758            & 140.6530  & 45.0216  & 0.333       & 5    \\
22425            & 274.9627  & 57.1797  & 0.166       & 5    \\
22576            & 131.5407  & 1.6167   & 0.260       & 5    \\
22637            & 117.2192  & 59.6907  & 0.410       & 5    \\
22918            & 57.0388   & 24.2672  & 0.109       & 5    \\
23463            & 264.0015	 & 67.9892  & 0.118       & 4    \\
24761            & 271.7904  & 59.2496  & 0.222       & 5    \\
24762            & 272.0796  & 59.2580  & 0.100       & 4    \\
\noalign{\smallskip}\hline
\end{tabular}
\tablefoot{RA and DEC are given in degrees. "S" is the redshift quality flag described in the present Section. The table is available at the CDS.}
\end{table}

\begin{figure*}
    \centering
    \includegraphics[width=\linewidth]{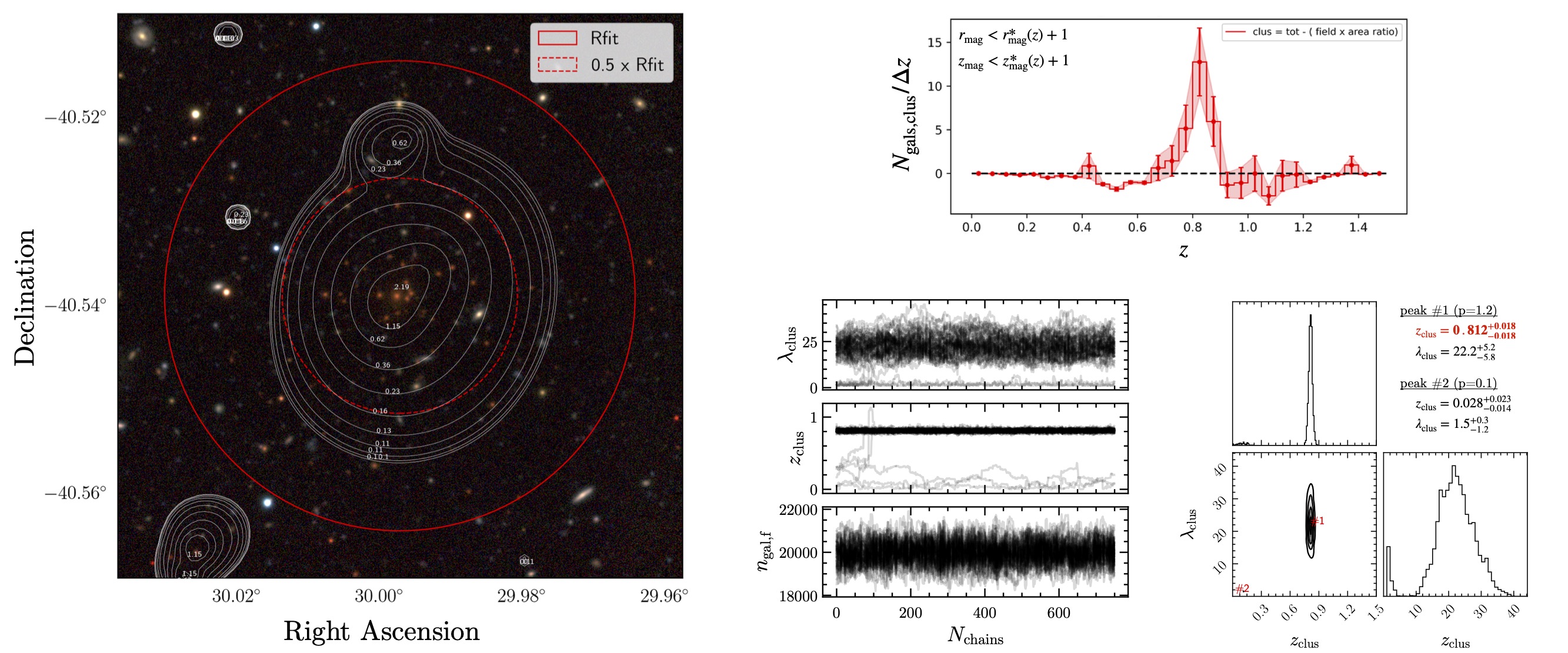}
    \caption{Example of the data available for photometric redshift inspection and validation with {\tt photXclus}. xclass22690 was identified at a photometric redshift of 0.81 after inspection of the $grz$ 3-color image from the LS DR9, with X-CLASS X-ray contours overplotted in white (Left). The histogram of the background-subtracted number of galaxies in bins of photometric redshifts is shown on the top right, while the diagnosis plots from {\tt photXclus} are shown on the bottom right (Left: MCMC traces; Right: corner plot of the marginal posteriors for $z_{\rm clus}$ and $\lambda_{\rm clus}$).
    }
    \label{fig:inspection}
\end{figure*}

We have defined a series of flags aggregated in the "ZSP" quality flag, with all definitions given below.
The first component of the flag ("Z") corresponds to the way the new redshift is attributed to each cluster. 

\begin{itemize}
    \item Z=0: no redshift
    \item Z=1: spec-z redshift with discrepant or no photo-z
    \item Z=2: spec-z redshift with consistent photo-z
    \item Z=3: photo-z redshift with discrepant tentative spec-z or no spec-z
   \item Z=4: photo-z redshift with consistent tentative spec-z.
\end{itemize}
If a spectroscopic redshift for the cluster is available (Z = 1 or 2), it is generally preferred over the photometric redshift. For clusters with Z = 3, the final redshift is set to the photometric redshift value as well as in cases where only one spec-z is available (Z = 4). For those with no redshift available, they are given Z = 0 and a global flag 000 (69 clusters). 

\smallskip
"S" is a flag that corresponds to details relating to the spectroscopic redshift status, with the following definitions: 
\begin{itemize}
    \item S=0: no spectroscopic redshift
    \item S=1: 1 spec-z in the cluster
    \item S=2: 1 spec-z, supposedly the BCG
    \item S=3: 2 spec-z in the cluster
    \item S=4: 2 spec-z, one supposedly the BCG
    \item S=5: 3 or more spec-z in the cluster.
\end{itemize}
"P" is a flag that corresponds to the photometric redshift status with the following criteria:
\begin{itemize}
    \item P=0: no photometric redshift
    \item P=1: not checked visually
    \item P=2: unclear
    \item P=3: possible
    \item P=4: validated
\end{itemize}
Special attention was paid to the 281 sources that did not have a redshift (neither photometric nor spectroscopic) in the previous version of the X-CLASS database \citepalias{2021Koulouridis}, in order to introduce a reliable and verified value. 
A visual inspection of a 3-color $grz$ image centered on the cluster was carried out by a team of reviewers. The photometric redshift histograms in the cluster and control regions as well as the {\tt photXclus} posterior distribution were added for the analysis (Fig.~\ref{fig:inspection}). Flags P = 2, 3, or 4 were assigned after consultation between the various evaluators, depending on the convergence of their classification. 

Finally, we grouped all the above ZSP flag combinations into three categories, based on the status initiated by \citetalias{2021Koulouridis}. These three classes are broader and essentially separate clusters that are spectroscopically confirmed and those that have only photometric redshift estimates. In the previous version of the catalogue, 87 sources were classified as Provisional. Each of these sources was visually inspected by a team of reviewers. The inspection considered the X-ray and optical images and any available spectroscopic or photometric data for galaxies in the region. Of the 87 sources, 47 were removed from the catalog, while 23 were assigned either a 000 flag or another appropriate flag if new information was available. The remaining 17 sources, which received mixed assessments, were also retained in the catalogue.

We present in Table~\ref{table:fullredshifts} the entire table of updated X-CLASS redshifts. The column 'flag' refers to ZSP as described in this section. For example, a cluster with a flag 304 has a validated photometric redshift but no spectroscopic information; therefore, it has a Photometric status. A flag 122 indicates that there is a spectroscopically confirmed BCG and that the photometric redshift of the cluster is questionable; but it has a Confirmed status. Finally, a flag 233 indicates a cluster with a spectroscopically confirmed BCG that is consistent with the photometric redshift. In this case, the new redshift is confirmed. 

\begin{table}[ht]
\caption{The 1599 X-CLASS galaxy clusters.}
\label{table:fullredshifts}
\centering 
\begin{tabular}{r r r c c l}
\hline\hline\noalign{\smallskip}
xclass & RA & DEC & $z$ & ZSP & status \\
 & (deg) & (deg) & & (flag) & \\
\noalign{\smallskip}\hline\noalign{\smallskip}
0020  &	193.4380  &	10.1954  &	0.654 & 221 & confirmed  \\
0023 & 194.2860 & -17.4119 &  0.047 & 251 & confirmed  \\
0033 & 193.6790 & -29.2227 &  0.056 & 251 & confirmed  \\
0034 & 193.5950 & -29.0162 & 0.053 & 251 & confirmed  \\
0035 & 196.2740 & -10.2802 & 0.330 & 301 & photometric \\
0038 &  36.5674  & -2.6651 & 0.056 & 251 & confirmed  \\
0039 &  36.4987 &  -2.8272 & 0.281 & 251 & confirmed  \\
0040 &  35.1871 & -3.4339 & 0.327 & 251 & confirmed  \\
0042 & 150.1230 & -19.6282 & 0.470 & 251 & confirmed \\
... &   ... &   ... &   ... &   ...  &  ...  \\
\noalign{\smallskip}\hline
\end{tabular}
\tablefoot{Each cluster are given with updated redshifts ($z$), associated flags and confirmation status. Only part of the table is presented here. The full table is available at the CDS.}
\end{table}

\section{Results of X-ray measurements\label{sec:measurements}}
Table~\ref{table:measurements} shows the complete X-ray measurement results for the X-CLASS-LR sample for clusters where the fits converged.

\begin{table*}
    \caption{List of the 155 galaxy clusters constituting the X-CLASS-LR sample. \label{table:measurements}}
    \centering
    \begin{tabular}{lllllllll}
    \hline\hline\noalign{\smallskip}
    X-CLASS & RA & DEC & $z$ & T & Z &  $L_X$  & $R_{500}$ & $\chi^2/\nu$ \\
    & ($\deg$) & ($\deg$) & & (keV) & ($Z_{\odot}$) &  ($10^{42}$ erg/s) & (kpc) \\
    \noalign{\smallskip}\hline\noalign{\smallskip}
0054   & 145.9380 &	16.7381 & 0.180 & $2.0^{+0.4}_{-0.3}$ & $0.2^{+0.2}_{-0.1}$          & $6.3^{+0.4}_{-0.5}$ & 648 &    1        \\ \\
0083   & 148.4240 &	1.6995  & 0.097 & $2.4^{+0.6}_{-0.4}$ & $1.0^{+0.7}_{-0.4}$          & $20^{+0.5}_{-0.5}$ & 744  &     68     \\ \\
0103  & 28.3062 &	0.8846  & 0.132 & $0.9^{+0.3}_{-0.2}$ & 0.3 *        & $2.0^{+0.5}_{-0.4}$ & 403 &      2     \\ \\
0108  & 10.8933 &	1.0163  & 0.195 & $1.7^{+0.7}_{-0.3}$ & $0.9^{+2.0}_{-0.5}$          & $10^{+1.0}_{-0.9}$ & 595  &     2     \\ 
...  & ... & ... & ... & ... & ... & ... & ...  & ... \\
\noalign{\smallskip}\hline
\end{tabular}
\tablefoot{For each cluster, we give the name (1), coordinates (2,3), redshift (4), measured temperature (5), measured metallicity (*:~fixed metallicity) (6), X-ray luminosity in the [0.5-2] keV band within $R_{500}$ (7), estimated radius (no uncertainties provided) (8), and the reduced chi-square of the pointing background level fit (9). A few measurements are not available (see Table~\ref{tab:selection}), and they appear as blank fields in the table. Only part of the table is presented here. The full table is available at the CDS.}
\end{table*}

\end{appendix}

\end{document}